\documentclass[preprint]{elsarticle} % default for submission
\usepackage{pgf}
\usepackage{lineno,hyperref}
\modulolinenumbers[5]

\usepackage{xcolor}
\definecolor{revision1}{rgb}{0,0,0} 

\journal{Nuclear Instruments and Methods in Physics Research, Section A }

\begin{document}

\begin{frontmatter}

\title{A Magnetic Field Cloak For Charged Particle Beams}

%\author{Name \fnref{myfootnote}}
%\address{Stony Brook University}

%\author{R. Cervantes,  K. Dehmelt, A. Deshpande, N. Feege, T. K. Hemmick, (SBU~students)}

% authors with major contribution
\author[sbu,bnlm]{K.~G.~Capobianco-Hogan}
\author[sbu,uwa]{R.~Cervantes}
\author[sbu,bnlp]{A.~Deshpande}
\author[sbu]{N.~Feege\corref{mycorrespondingauthor}}
\ead{nils.feege@stonybrook.edu}
\author[sbu,uva]{T.~Krahulik}
\author[sbu,uwa]{J.~LaBounty}
\author[sbu]{R.~Sekelsky}

% additional authors
\author[sbu]{A.~Adhyatman}
\author[sbu]{G.~Arrowsmith-Kron}
\author[sbu,bnl]{B.~Coe}
\author[sbu]{K.~Dehmelt}
\author[sbu]{T.~K.~Hemmick}
\author[sbu]{S.~Jeffas}
\author[sbu]{T.~LaByer}
\author[sbu]{S.~Mahmud}
\author[sbu]{A.~Oliveira}
\author[sbu]{A.~Quadri}
\author[sbu]{K.~Sharma}
\author[sbu,neu]{A.~Tishelman-Charny}

\address[sbu]{Department of Physics and Astronomy, Stony Brook University, Stony Brook, NY 11794, USA}
\address[bnlm]{Superconducting Magnet Division, Brookhaven National Laboratory, Upton, NY 11973, USA}
\address[bnlp]{Physics Department, Brookhaven National Laboratory, Upton, NY 11973, USA}

\address[uwa]{Department of Physics, University of Washington, Seattle, WA 98195, USA}
\address[uva]{Department of Physics, University of Virginia, Charlottesville, VA 22904, USA}
\address[bnl]{Collider Accelerator Department, Brookhaven National Laboratory, Upton, NY 11973, USA}
\address[neu]{Physics Department, Northeastern University, Boston, MA 02115, USA}

\cortext[mycorrespondingauthor]{Corresponding author}



\begin{abstract}
%Concise + factual. State purpose of research, principle results, and major conclusions. Problem? So what? Solution? Benefits? Inspire! 
Shielding charged particle beams from transverse magnetic fields is a common challenge for particle accelerators and experiments. We demonstrate that a magnetic field cloak is a viable solution. It allows for the use of dipole magnets in the forward regions of experiments at an Electron Ion Collider (EIC) and other facilities without interfering with the incoming beams. The dipoles can improve the momentum measurements of charged final state particles at angles close to the beam line and therefore increase the physics reach of {\color{revision1}these} experiments. In contrast to other magnetic shielding options (such as active coils), a cloak requires no external powering. We discuss the design parameters, fabrication, and limitations of a magnetic field cloak and demonstrate that cylinders made from 45 layers of YBCO high-temperature superconductor, combined with a ferromagnetic shell made from epoxy and stainless steel powder, shield more than 99\% of a transverse magnetic field of up to 0.45~T (95~\% shielding at 0.5~T) at liquid nitrogen temperature. The ferromagnetic shell reduces field distortions caused by the superconductor alone by 90\% at 0.45~T.
%In contrast to other magnetic shielding options (such as active coils), a cloak requires no external powering.
\end{abstract}


\end{frontmatter}

%\linenumbers

%______________________________________________________________________________
%INTRODUCTION: Objectives of the work, adequate background, NO summary of results!
\section{Introduction}

\begin{figure}
	\centering
	\includegraphics[width = \textwidth]{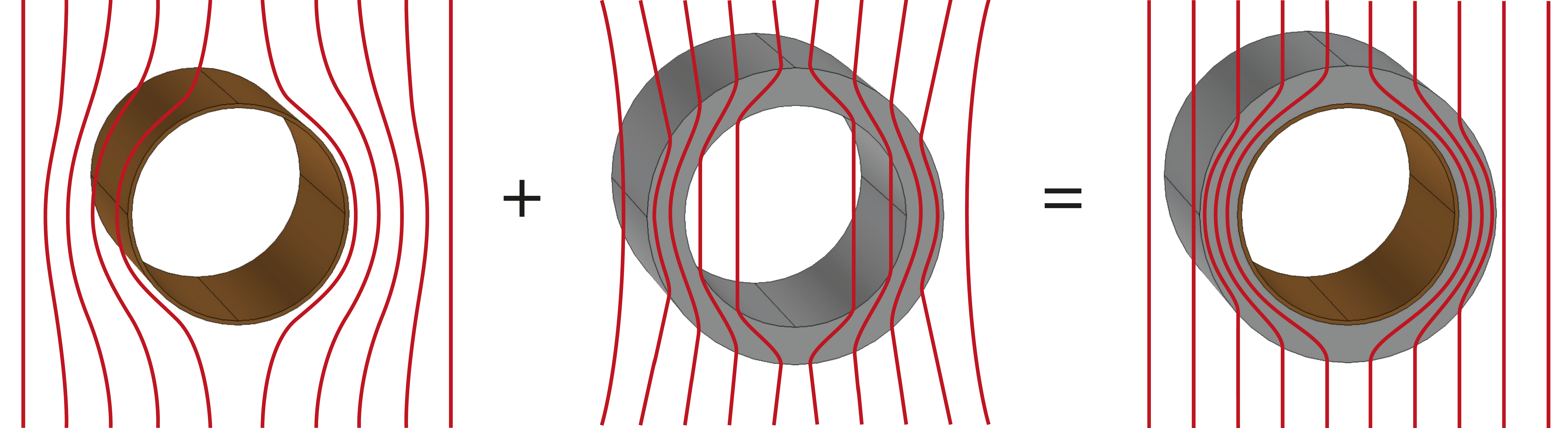}
    	\caption{Concept of a magnetic field cloak. From left to right: A superconducting cylinder pushes out magnetic field lines, a ferromagnetic cylinder pulls in magnetic field lines, and the combination of both  forms a cloak (given the correct thickness and magnetic permeability of the ferromagnet).}
    	\label{fig:conceptual_cloak}
\end{figure}
Magnetic fields are routinely used to steer charged particle beams and to analyze the momenta of charged particles produced in fixed-target and collider experiments. The field component transverse to the trajectory of a charged particle deflects it and, in the case of a polarized beam crossing a field gradient, depolarizes it. Beams at particle collider facilities need adequate shielding from fields that would cause disturbances. Established designs of magnetic field shields use cylinders made from low-temperature superconductors \cite{Martin:1972xd}. Magnetic flux lines incident on a superconducting cylinder induce screening currents, and the magnetic fields generated by these currents counteract the external field. As a result, the inside of the cylinder remains field-free, while the field on the outside is distorted. This distortion can be corrected by adding a ferromagnetic shell around the superconductor. Unlike the superconductor, a ferromagnetic shell {\color{revision1}pulls in magnetic flux lines. The combination} of superconductor and ferromagnet forms a magnetic field cloak (see Fig.~\ref{fig:conceptual_cloak}). The ferromagnet of a superconductor-ferromagnet bilayer effectively contains all field distortions caused by the superconductor if its magnetic permeability $\mu_r$ is tuned to
\begin{eqnarray}
    \mu_r = \frac{R_2^2 + R_1^2}{R_2^2 -R_1^2},   
    \label{eqn:permeability}
\end{eqnarray}
where $R_1$ and $R_2$ are the inner and outer radius of the ferromagnet ($R_1$ is also the outer radius of the superconductor)~\cite{gomory2012}. Thus, a cloak can provide a field-free tunnel without disturbing the external field.

Magnetic field cloaks are topics of active research \cite{Giunchi:2016ncd,3dcloak2015}. We want to demonstrate that our design, which uses high-temperature superconductor (HTS) cylinders, is a viable solution to cloak charged particle beams at future particle accelerator facilities such as the Electron Ion Collider (EIC). Such a facility would require a cloak that shields a magnetic field of at least 0.5~T over a length of 1~m. Section~\ref{sec:scshielding} briefly summarizes the basics of shielding magnetic fields with superconductors, Sec.~\ref{sec:prototypes} explains the fabrication of our superconductor shields and cloak prototypes, Sec.~\ref{sec:setups} describes our test setups, Sec.~\ref{sec:results} presents the results of magnetic field shielding and cloaking measurements with our prototypes, and Sec.~\ref{sec:conclusion} gives our conclusions.

%______________________________________________________________________________
\section {Shielding magnetic fields with high temperature superconductors}
\label{sec:scshielding}
The response of a type-II superconductor to magnetic fields is characterized by two critical fields: $B_{c1}$ and $B_{c2}$. If a cylinder made from such a material is exposed to a transverse magnetic field below $B_{c1}$, the flux lines are bent around the cylinder. Above a certain threshold field, the field penetrating the superconductor has an approximately logarithmic time dependence~\cite{RevModPhys.68.911}. Between $B_{c1}$ and $B_{c2}$, flux vortices form and allow the field to partially seep through the superconductor. In this field range, stacking multiple layers of superconductor improves the overall magnetic field shielding. Above $B_{c2}$, the superconductivity is destroyed and the field passes through the shield. The critical fields of a superconductor become larger the lower the superconductor's temperature falls below its critical temperature. %depend on its temperature and%, the higher the effective critical fields rise. %Therefore, cooling high-temperature superconductors far below their critical temperatures improves their shielding performance.

High-temperature superconductors (HTS) are especially convenient because they can be cooled below their critical temperature with liquid nitrogen (rather than liquid helium), resulting in relatively modest cryogenic costs and fast prototyping. Although various experiments have characterized HTS shields~\cite{Denis2007,Fagnard2009, Kvitkovic2011, Karthikeyan1994, Fagnard2010, Kvitkovic2009}, we are not aware of any substantial shielding past 10~mT with such materials, especially for transverse fields. Ref.~\cite{Kvitkovic2009} comes closest, but the authors only shield approximately 60\% of the field with an applied field of amplitude 50~mT and frequency 10~Hz. If HTS can be demonstrated to shield fields above 0.5~T, they can replace low temperature superconductor shields (such as NbTi sheets~\cite{nippon2002}).
% which are no longer being produced

%______________________________________________________________________________
\section{Prototype construction}
\label{sec:prototypes}

\subsection{1~m long / 2-layer and a 4.5~inch long / 4-layer HTS shield}
\label{sec:shieldbnl}

We use 46~mm wide superconductor wire insert manufactured by American Superconductor Inc (\url{www.amsc.com}) to fabricate superconducting cylinders. This width is an intermediate stage on their production line and, due to low demand, not currently supported as a product~\cite{Rupich2010}. The insert is made from a YBCO ceramic with a critical temperature of about 90~K. The ceramic is deposited on an oxide-buffered Ni-W alloy substrate and coated with silver. The width and flexibility of this superconductor allow us to combine two strips to a cylinder of up to approximately 1~inch diameter. The orientation of the strips along the cylinder axis facilitates the induction of supercurrents that generate a magnetic dipole field. Therefore, it is most effective for shielding transverse magnetic fields~\cite{nouri2013}. The maximum field that a cylinder can shield increases with the number of layers as long as the field does not exceed the second critical field of the superconductor~\cite{cervantes2015}. %(despite the break along the middle) %cosine-theta dipole magnet
\begin{figure}
	\centering
    	\includegraphics[width = 0.75\textwidth]{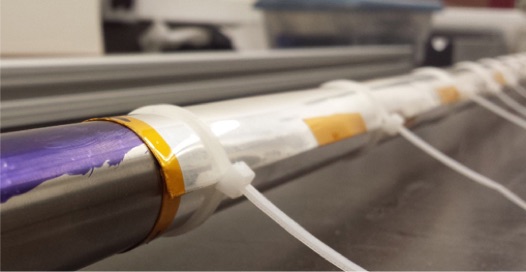}
    	\caption{2-layer HTS shield made from four 1~m long layers of 46~mm wide superconductor strips attached to a 60~inch stainless steel tube with 1~inch outer diameter. The strips form a double layer around the top and the bottom of the tube.}
    	\label{fig:sc_bnl}
\end{figure}
Figure~\ref{fig:sc_bnl} shows our HTS shield prototype consisting of four 1~m long superconductor strips attached to a 60~inch long stainless steel tube with 1~inch outer diameter. The superconductor strips form a double layer on the top and the bottom of the pipe and overlap at the connecting sides. We use Kapton$\textsuperscript{\textregistered}$ tape and zip ties to hold the superconductor strips in place. In addition, we test a 4.5~inch long HTS shield prototype with four layers of superconductor strips attached to an aluminum tube.

\subsection{4.5~inch long / 45-layer HTS shield}
\label{sec:shield45}
\begin{figure}
	\centering
    	\includegraphics[height = 0.45\textwidth]{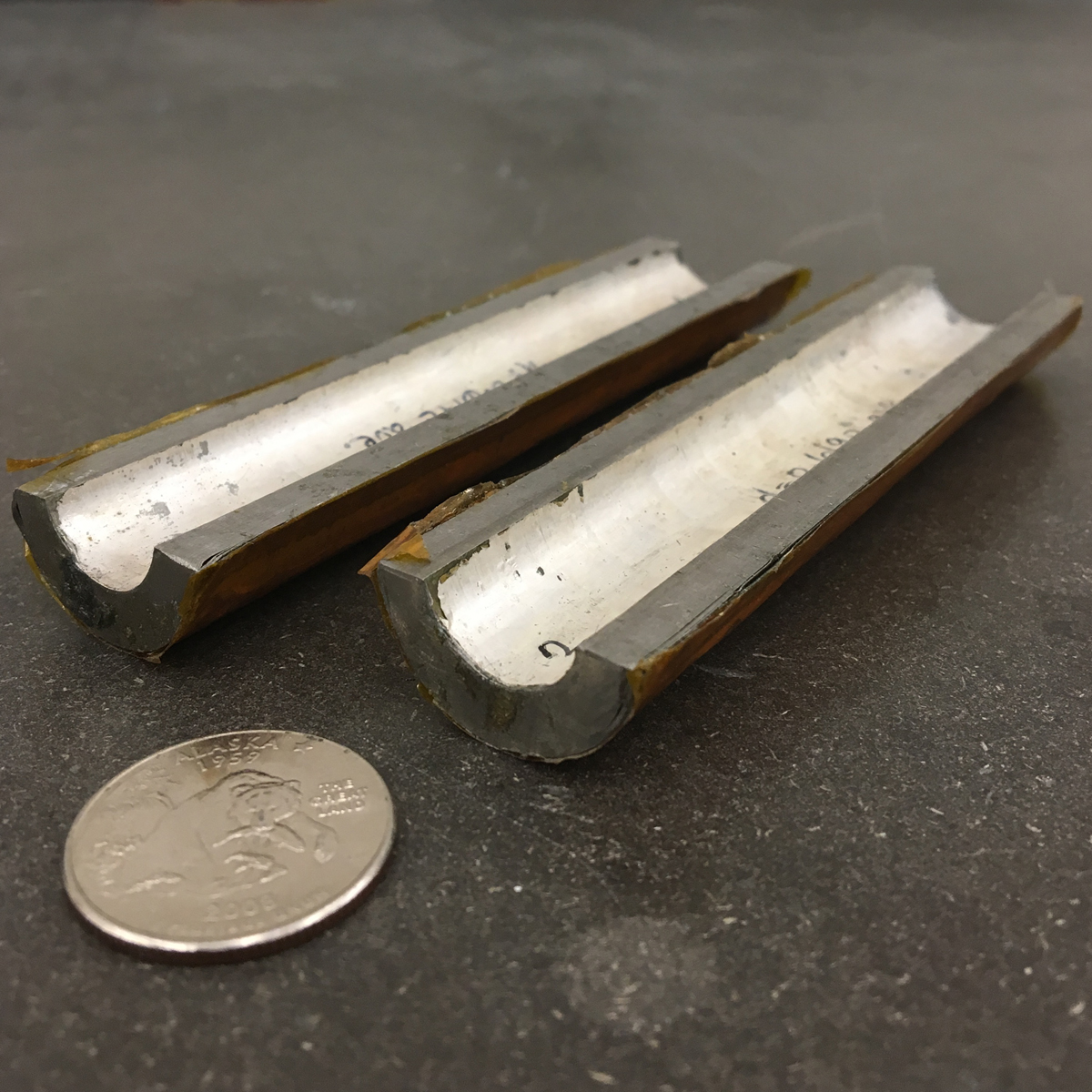}
    	\includegraphics[height = 0.45\textwidth]{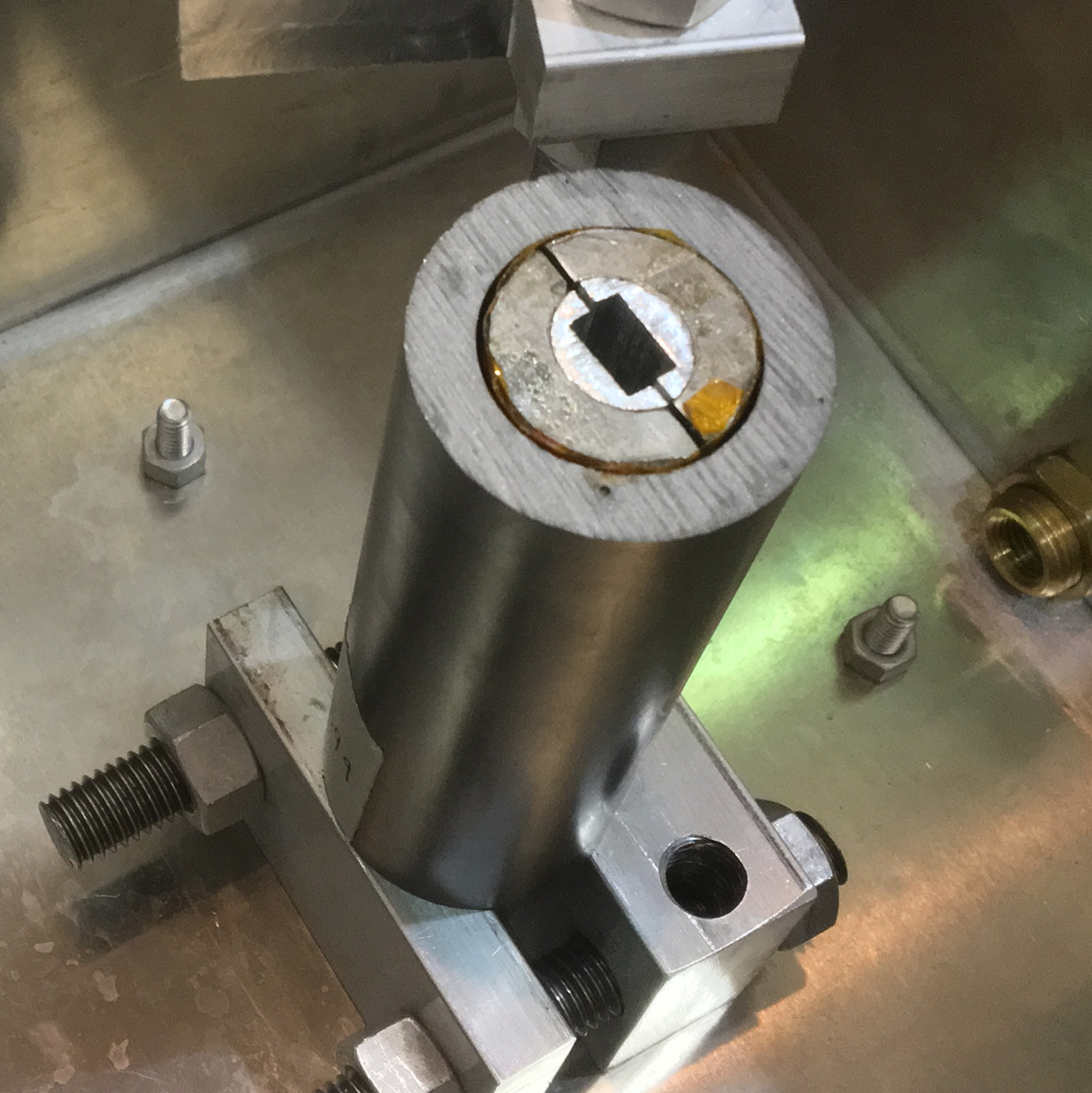}
    	\caption{Left panel: Two half-shells of a 4.5~inch long, 1~inch outer diameter, 45-layer HTS shield. Right panel: Fully assembled cloak with the 45-layer HTS shield and a ferromagnetic epoxy / steel powder shell. An aluminum core holds the cloak in place.}
    	\label{fig:sc_45layers}
\end{figure}

Using a fabrication process based on \cite{Martin:1972xd}, we build a HTS shield with 45 layers of 46~mm wide American Superconductor HTS wire insert. The process uses a die-and-mandrel setup heated in an oven to laminate multiple layers of superconductor wire and solder. Removing excess superconductor and solder on the sides with a milling machine creates half-cylinders. We combine two of these half cylinders to form a full shielding tube. The left panel of Fig. ~\ref{fig:sc_45layers} shows the two halves of our 4.5~inch long, 1~inch outer diameter HTS shield with 45 layers. {\color{revision1}The thickness of this shield varies between 0.22~inch and 0.26~inch along its circumference.} Based on an extrapolation method described in~\cite{cervantes2015} and measurements for single layer shielding, we expect this 45-layer prototype to shield more than 99\% of a transverse magnetic field up to 0.5~T.

\subsection{4.5~inch long / 4-layer and 4.5~inch long / 45-layer HTS cloak}
\label{sec:cloak45}
To fabricate a ferromagnetic shell, we mix 430 stainless steel powder (magnetic permeability $\mu \approx 1000$~\cite{oxley2009}) with commercial epoxy and pour the mixture into a tubular mold. {\color{revision1} We keep the mold upright to help air bubbles accumulate at the top and invert it every minute for 30~minutes to prevent the steel powder from setting while the epoxy is hardening. When placing the hardened cylinder in a 30~mT homogenous magnetic field perpendicular to its axis, we observe maximum field shielding at the center of the cylinder and a symmetric shielding profile around it (Fig.~\ref{fig:sbu_bvz} illustrates the same kind of shielding measurement performed on superconducting cylinders). This confirms the uniform distribution of steel powder in the ferromagnetic shell.} Controlling the amount of stainless steel powder allows us to control the permeability of the mixture~\cite{rozanov2008} (see Sec.~\ref{sec:fmtuning} for more details). {\color{revision1} We use a $4.5$~inch long ferromagnetic shell with inner radius $R_1 =  0.533(2)$~inch and outer radius $R_2 = 0.805(1)$~inch to assemble cloak prototypes with the 4-layer and 45-layer HTS shields. Given these dimensions, Eqn.~\ref{eqn:permeability} predicts that a magnetic permeability of $\mu_r \; = \; 2.56(2)$ yields perfect cloaking.} The right panel of Fig. \ref{fig:sc_45layers} shows our fully assembled magnetic field cloak prototype using the 45-layer HTS shield.

% Most commercial materials have  either $\mu_r = 1$ or $\mu_r > 100$. However, a ferromagnetic layer with an easily manufacturable thickness requires a permeability between 3 and 20. For example, a ferromagnetic layer of R$_1$ = 1.25 cm and R$_2$ = 1.5 cm needs $\mu_r \approx 5.5$. 

%______________________________________________________________________________
\section{Test setups}
\label{sec:setups}

\subsection{The Van de Graaff Facility at Brookhaven National Laboratory}
\begin{figure}
    	\centering
    	\includegraphics[width=0.75\textwidth]{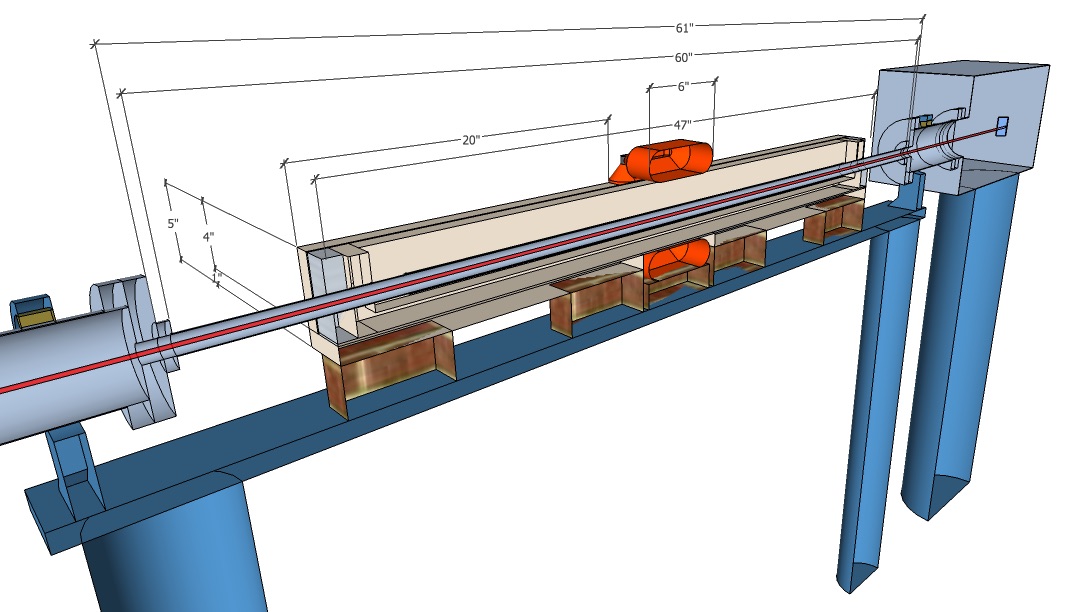}
    	\caption{The beam test setup in the BNL Van de Graaff beam line (beam entering from the left) showing the stainless steel tube with superconductor shield inside the aluminum box filled with liquid nitrogen, the steering dipole magnet in the center, and the target chamber with zinc sulfide screen on the right.}
    	\label{fig:bnlsetup}
\end{figure}

The Tandem Van de Graaff Facility at Brookhaven National Laboratory provides users with a variety of ion beams. {\color{revision1}We use the facility to test how well our 1~m HTS shield (see Sec.~\ref{sec:shieldbnl}) shields ion beams from a magnetic dipole field.} Figure \ref{fig:bnlsetup} shows a cross section of the test setup. The stainless steel core of our shield connects to the beam line, and a five-sided aluminum box insulated with 1~inch thick extruded polystyrene foam plates holds liquid nitrogen to cool the superconductor. A steering dipole magnet (a square arrangement of four coils with iron yokes, an inner opening of 4~inch by 4~inch, and field variations of less than 0.5~mT over a 1~inch diameter area in the center) is placed around the liquid nitrogen bath so that it creates a vertical magnetic field w.r.t. the beam line. We use Lakeshore's cryogenic transverse hall sensor (HGCT3020) and Model 425 Gaussmeter to measure the vertical component of the magnetic field in the center of the magnet as a function of the magnet current, as well as the field at different positions along the beam line at a fixed magnet current. In addition, we measure the deflection of $^{7}_{3}\mbox{Li}^{3+}$ and $^{16}_{8}\mbox{O}^{3+}$ ion beams as a function of the magnet current by recording the beam position with a zinc sulfide screen, which sits in a target chamber downstream of our HTS shield, and a digital camera. %While the Hall sensor measures the field at discrete points, the tests with ion beams are sensitive to the integrated field penetrating the HTS shield.
%  The screen  at a 45 degree angle. 
%  connected to a desktop computer
% The liquid nitrogen level is kept within a nominal range by an Arduino-based feedback loop, resistors inside the box that monitor the liquid nitrogen level, and a solenoidal valve that regulates the flow of liquid nitrogen from an external dewar to the cooling box.
%Conflat ends of the 60~inch stainless steel tube connect to flanges on the beam line

\subsection{Helmholtz coil setup}
\begin{figure}
	\centering
	\includegraphics[height = 0.3\textheight]{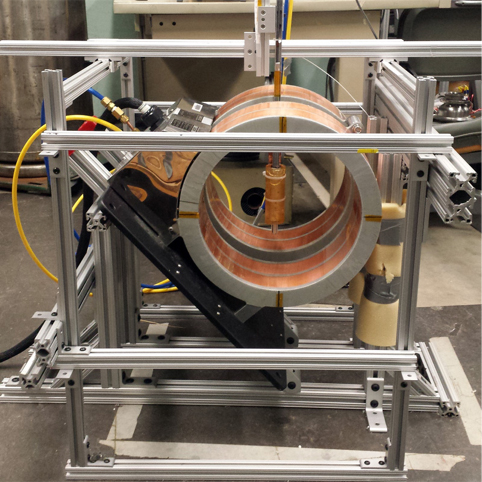}
    	\caption{The Helmholtz coil setup with our 80/20 rig. The rig holds our prototypes and allows us to move a Hall sensor in all three spatial directions inside and around the prototype under test.}
    	\label{fig:helmholtz_setup}
\end{figure}

We use the homogenous magnetic field generated by a pair of Helmholtz coils (GMW Model 5451, manufacturer quoted field uniformity $\Delta B/B$ less than $\pm$200ppm over a 30~mm sphere) to measure the permeability of ferromagnetic cylinders and to test magnetic field cloaking up to fields of 50~mT. A custom rig holds the prototype under test (superconductor cylinder, ferromagnetic cylinder, or full cloak) and a Hall sensor. The rig allows us to move the Hall sensor independently in three Cartesian directions to map the magnetic field inside of and around the prototype. We cool the prototypes by immersing them in a bath of liquid nitrogen confined in a box made from extruded polystyrene foam. Figure \ref{fig:helmholtz_setup} shows the setup.

%  a dipole with iron yoke (Systron Donner Model 7600, see Fig. \ref{fig:dipole_setup}) provides a very inhomogeneous but relatively strong field of up to $\approx$600~mT in the center.
% Figure \ref{fig:sc_experiment} illustrates how we test the magnetic field shielding performance of our superconductor shields: We place the cylinders in a magnet and measure the  field B$_i$ at the center of the cylinder as a function of the applied field B$_o$ 

\subsection{The 4 Tesla Magnet Facility at Argonne National Laboratory}
\begin{figure}
	\centering
	\includegraphics[width = 0.49\textwidth]{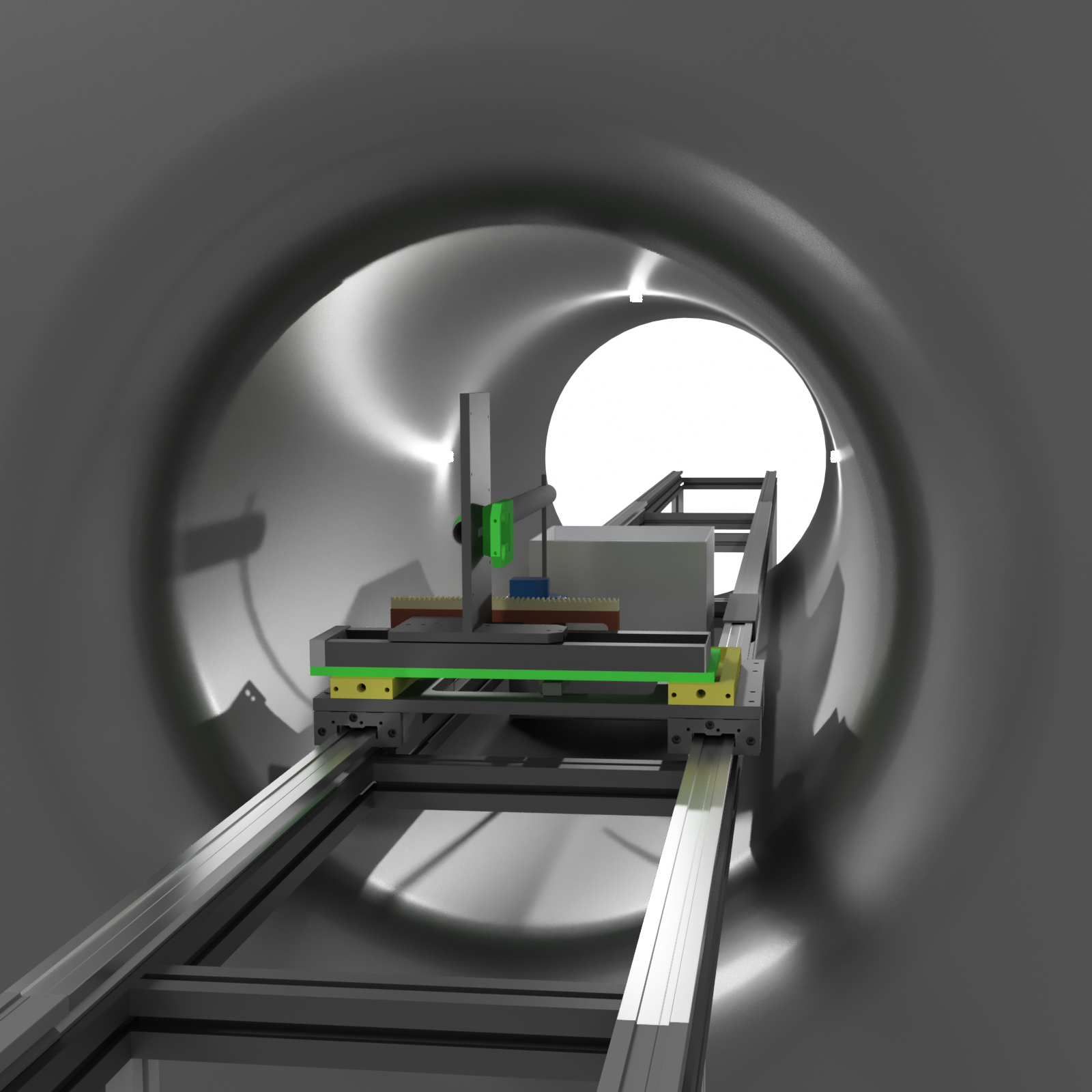}
	\includegraphics[width = 0.49\textwidth]{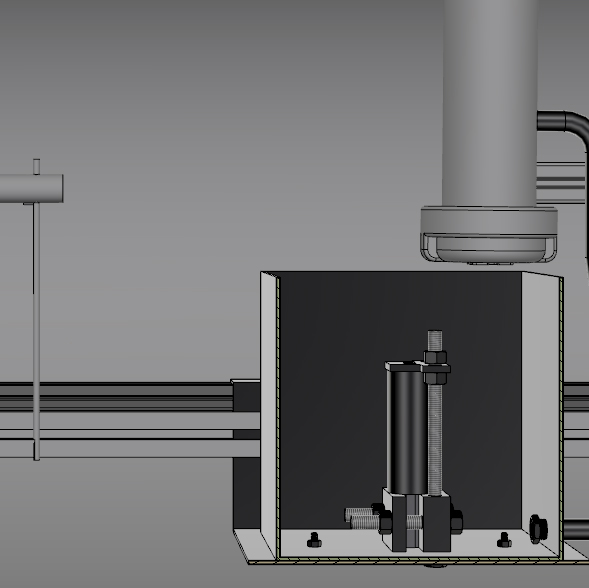}
    	\caption{Left panel: Experimental setup viewed from outside of the MRI magnet. Right Panel: Closeup view (without the MRI magnet) of the liquid nitrogen box supported by hanging aluminum plates, cloak prototype, Hall probe, probe holder, and translational stage to position the probe.}
    	\label{fig:mri_setup}
\end{figure}

The 4~Tesla Magnet Facility at Argonne National Laboratory gives users access to an MRI magnet to test detector components in a very homogenous field from 0 to 4~T. {\color{revision1}We use it to measure the permeability of ferromagnetic cylinders and to test magnetic field cloaking up to fields of 0.5~T.} Figure~\ref{fig:mri_setup} shows our test setup using this magnet. An open aluminum box (10.3~inch high, 10~inch wide, 10~inch deep) surrounded by 1~inch thick extruded polystyrene foam plates sits on a rail system inside the magnet and holds liquid nitrogen for cooling. Our 45-layer HTS shield, a ferromagnet cylinder, and a full cloak can be held  inside this box by a mount fixed to its base. We use a translational stage (three Cartesian degrees of freedom) on the rails to position a Hall sensor connected to a long rod to measure the magnetic field inside our prototypes and to map the magnetic field around them. % A tube is mounted vertically above the cooling box and acts as liquid-gas separator: A bulkhead connection on the side of the tube serves as input for the liquid-gas mixture. The tube is capped on the bottom, and an 0.5~inch hole is drilled in the cap. Liquid nitrogen percolates directly from this hole into our cooling box, while gas escapes out the top

%, including an external liquid nitrogen dewar

%different inserts for the different superconducting prototypes that we planto test. Each insert positions the superconductor 6 cm above the base and gives us enough room to measure the magnetic field 5 cm below the superconductor. The height of the box allows us to also measure the field 5 cm above the superconductor (accounting for the thickness of the base) without moving the Hall sensor out of the liquid nitrogen bath.

%______________________________________________________________________________
\section{Results}
\label{sec:results}
\subsection{Magnetic field shielding}

\begin{figure}
    	\centering
    	\includegraphics[width=0.75\textwidth]{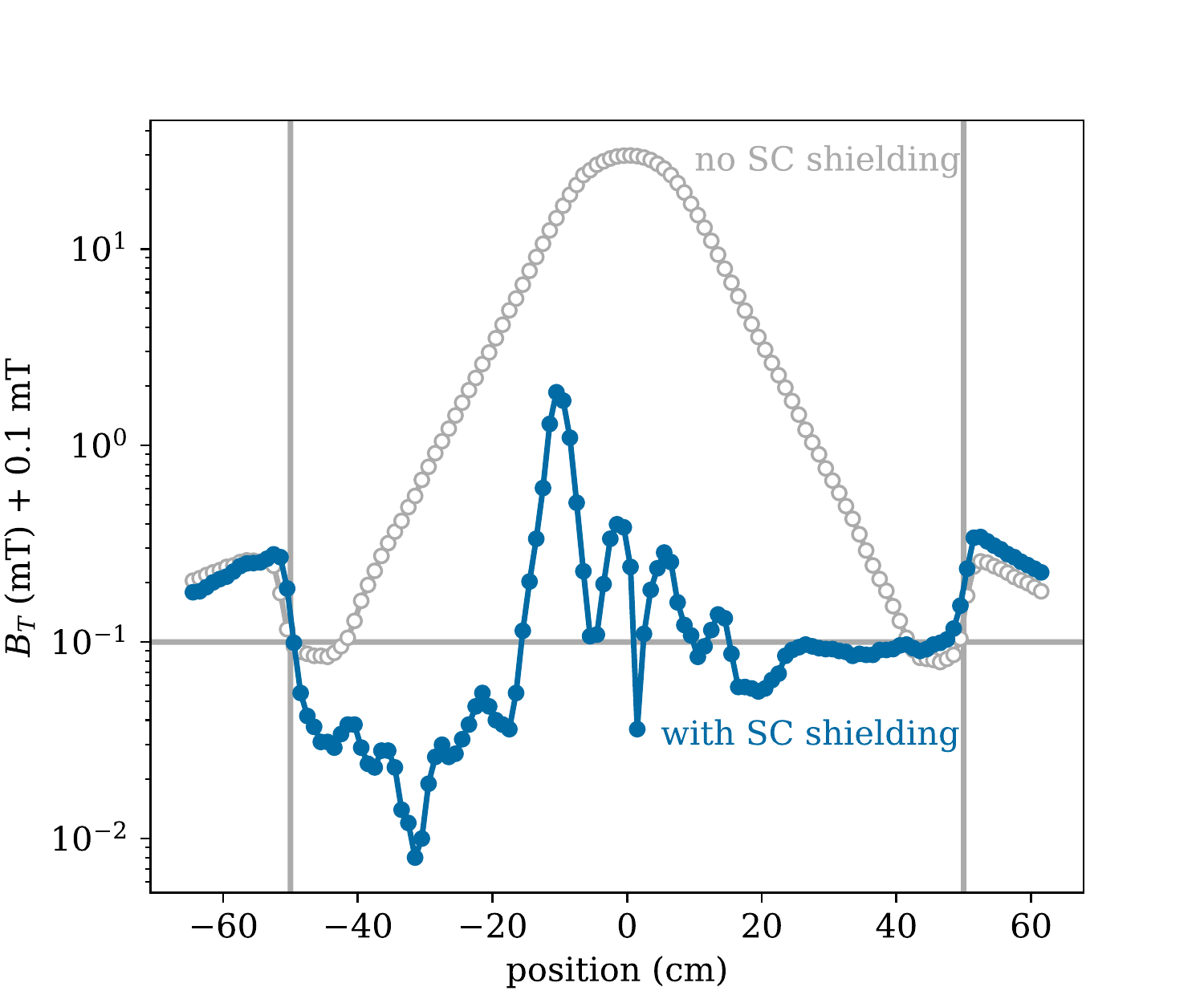}
    	\caption{Vertical component {\color{revision1}$B_T$} of the magnetic field measured in the Van de Graaff setup at different positions along the axis of the 1~m long, 2-layer HTS shield at room temperature (`no SC shielding`) and with liquid nitrogen cooling (`with SC shielding`) at a nominal steering dipole field of 30~mT. The vertical lines indicate the extension of the HTS shield. The ordinate uses logarithmic scale. In addition, an offset of 0.1~mT (indicated by the horizontal grey line) is added to each measurement.}
    	\label{fig:bnl_bvz}
\end{figure}

\begin{figure}
    	\centering
    	\includegraphics[width=0.75\textwidth]{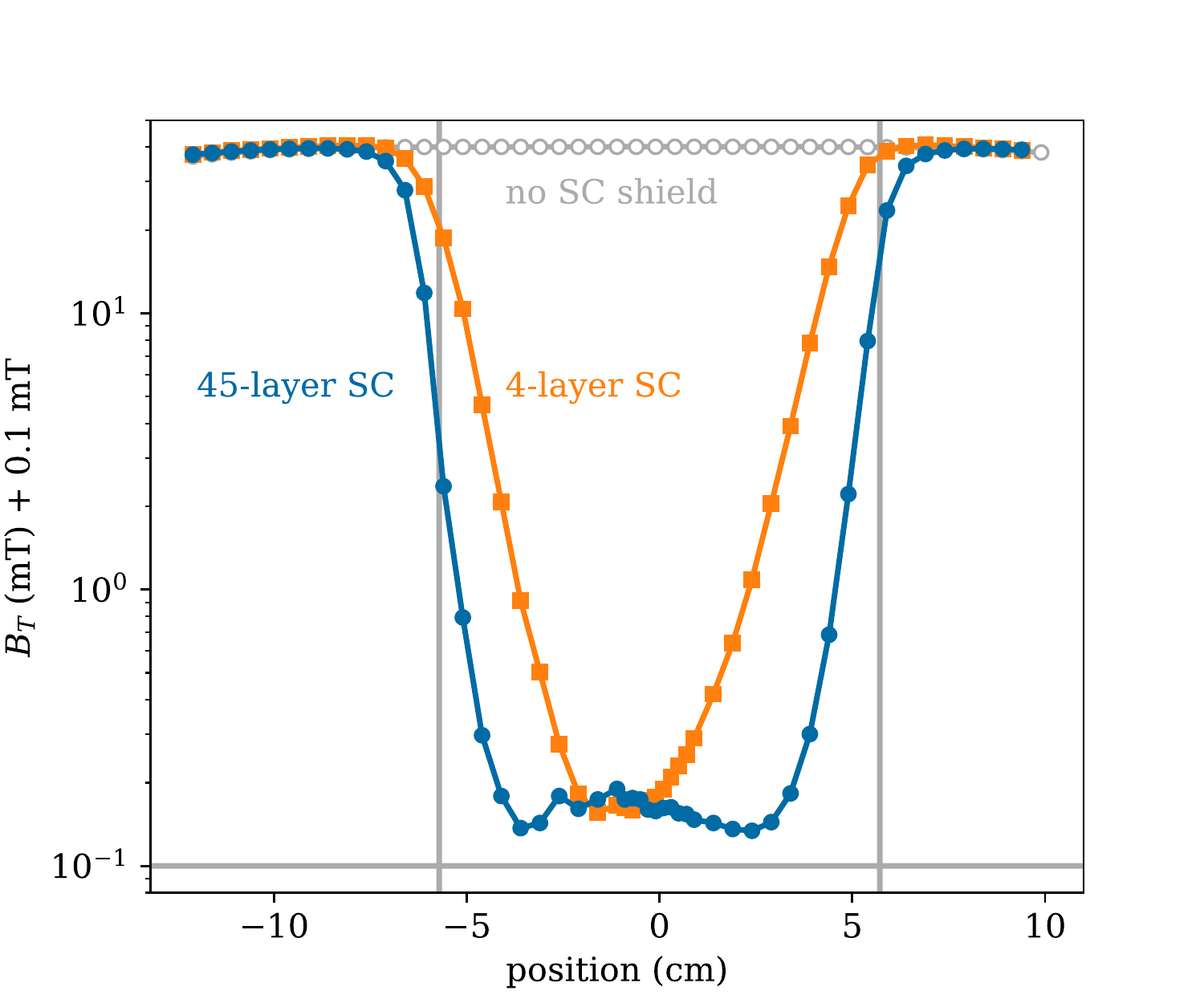}
    	\caption{Magnetic field component {\color{revision1}$B_T$} measured at various positions along the axis of the 4.5~inch long, 4-layer HTS and 4.5~inch long, 45-layer HTS shields in the Helmholtz coils setup at a nominal field of 40~mT. The vertical lines indicate the extension of the HTS shields. The ordinate uses logarithmic scale.}
    	\label{fig:sbu_bvz}
\end{figure}

\begin{figure}
 	\centering
    	\includegraphics[width = 0.75\textwidth]{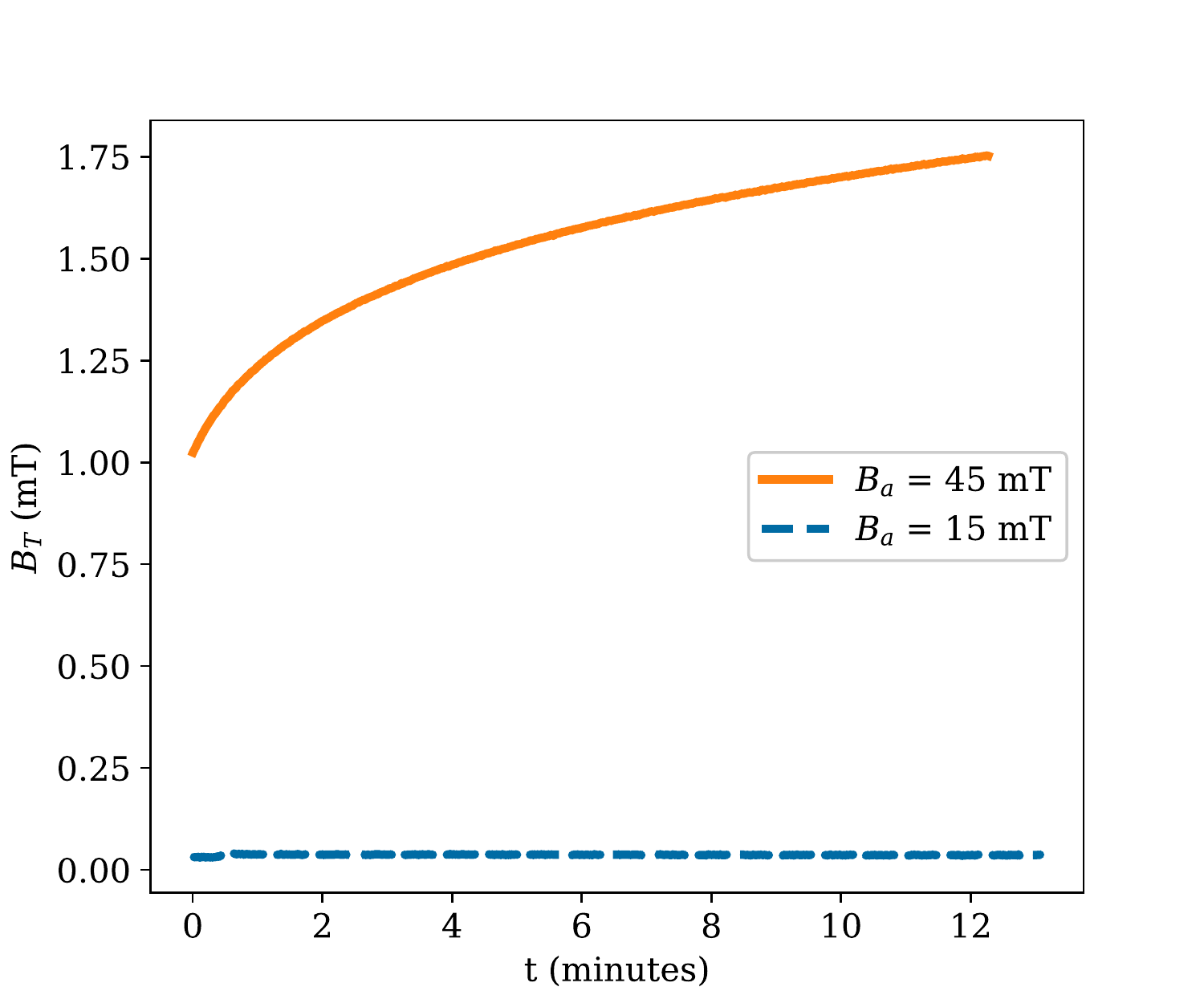}
    	\caption{Magnetic field component {\color{revision1}$B_T$} measured in the center of the 1~m long, 2-layer HTS shield in the Van de Graaff setup as a function of measurement time $t$ for nominal steering dipole magnet fields of {\color{revision1}$B_{a}$~=~15~mT and $B_{a}$~=~45~mT.}}
    	\label{fig:B_v_time}
\end{figure}

\begin{figure}
    	\centering
    	\includegraphics[width=0.75\textwidth]{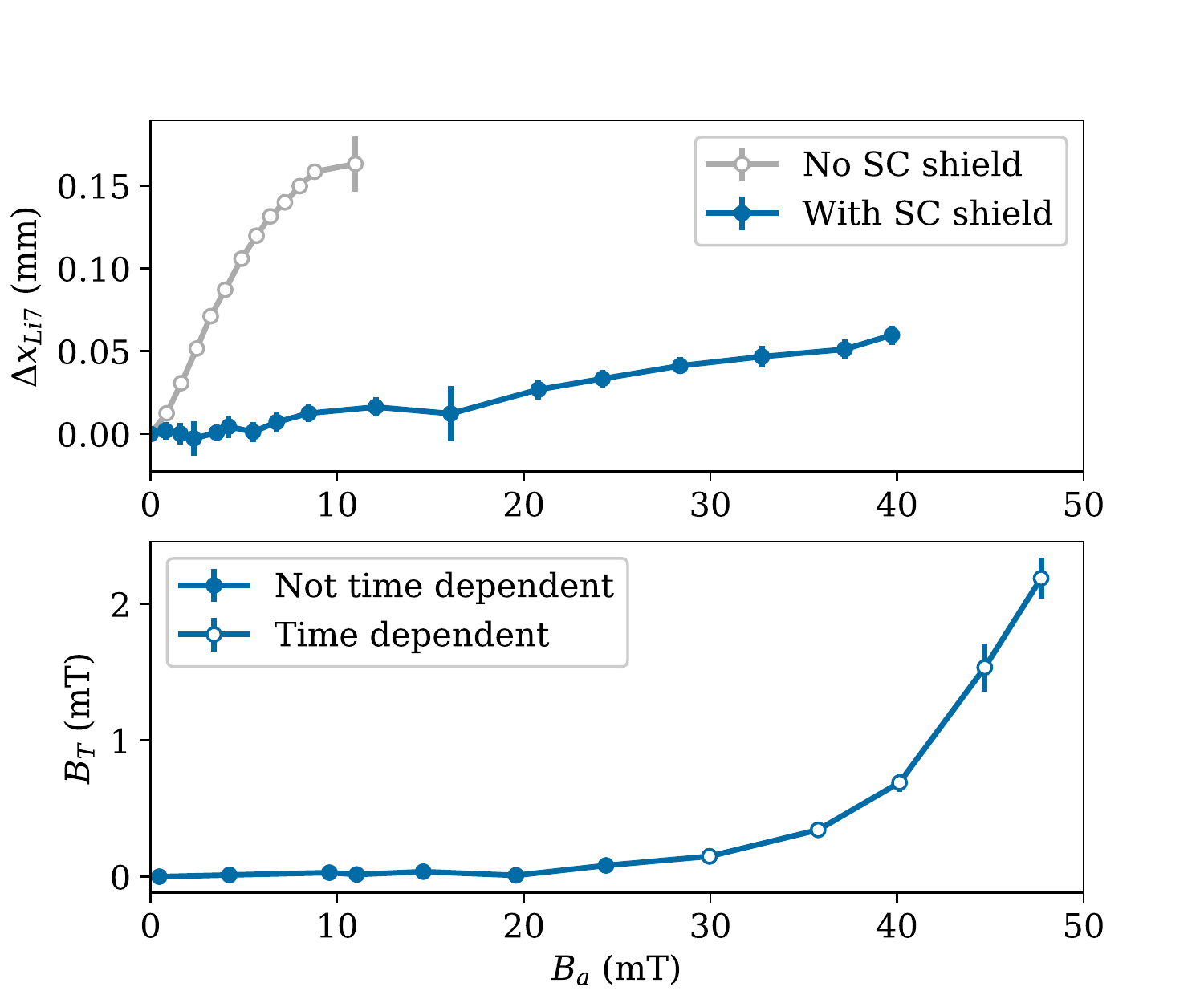}
    	\caption{Top panel: Displacement $\Delta x_{Li7}$ of the $^{7}_{3}\mbox{Li}^{3+}$ beam passing through the 1~m long, 2-layer superconductor shield  in the Van de Graaff setup (at room temperature and in its superconducting state) as a function of the nominal steering dipole field {\color{revision1}$B_{a}$}. Bottom panel: Mean field {\color{revision1}$B_T$} measured in the center of the shield as a function of the steering dipole field. Open symbols indicate time dependence of the measured field.}
    	\label{fig:bnl_bvi}
\end{figure}

\begin{figure}
    	\centering
    	\includegraphics[width = 0.75\textwidth]{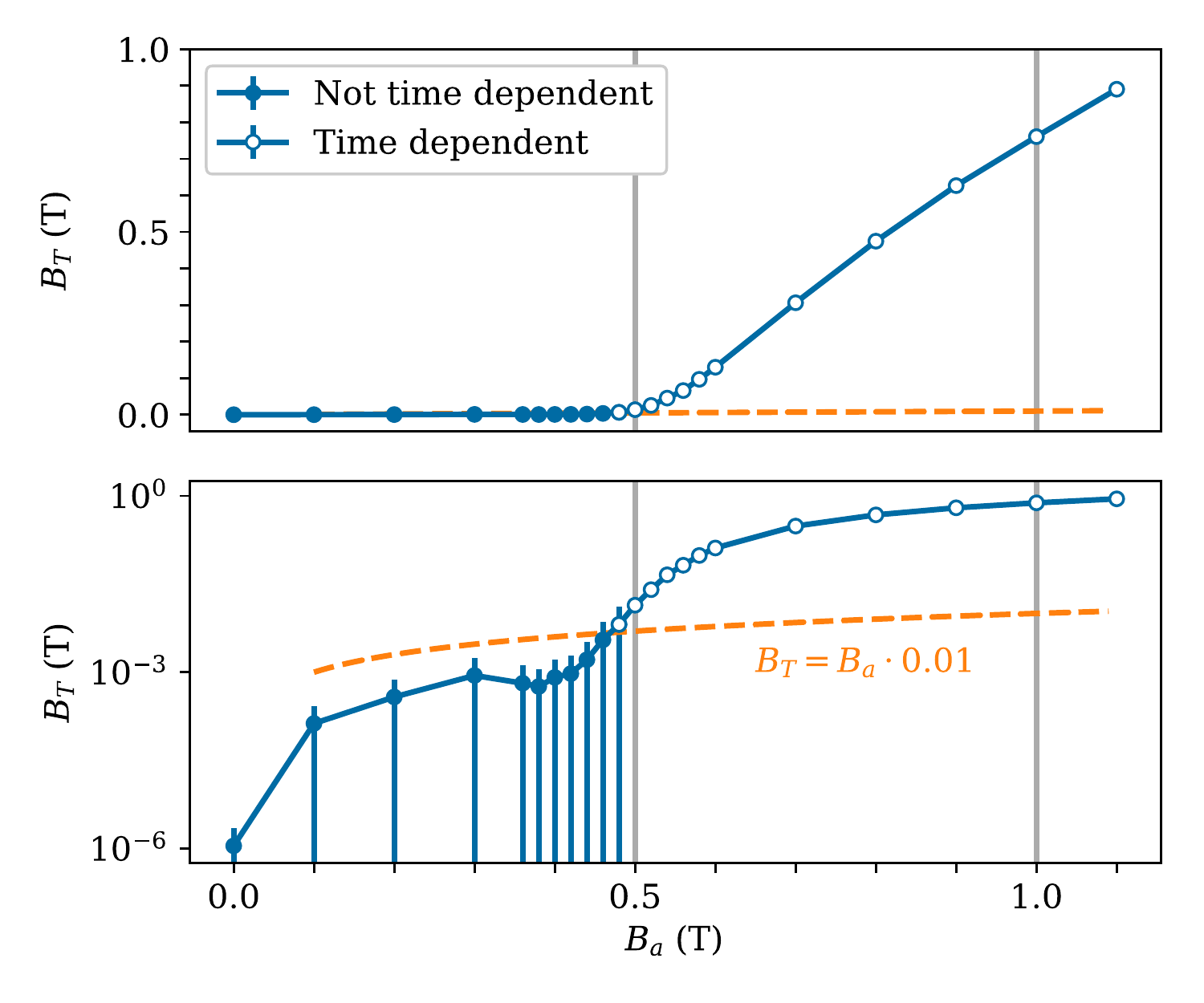}
    	\caption{Magnetic field component {\color{revision1}$B_T$} measured in the center of the 4.5~inch long, 45-layer HTS shield prototype inside the MRI magnet as a function of the nominal magnetic field $B_{a}$ in linear scale (top panel) and logarithmic scale (bottom panel). The open markers indicate field measurements showing an increase over time. The vertical lines mark $B_a = 0.5$~T and $B_a = 1.0$~T. {\color{revision1}A dashed line indicating $B_{T} = B_{a} \cdot 0.01$} is shown as well.}
    	\label{fig:sc_shield_mri}
\end{figure}

% SHIELDING
We use all three setups described in Sec.\ref{sec:setups} to test different aspects of shielding magnetic fields with our HTS shields. In the Van de Graaff setup, the 1~m long, 2-layer HTS shield extends significantly past the magnet. Therefore, we expect only minimal field leaking in through the ends of the shield. Figure \ref{fig:bnl_bvz} shows the vertical component {\color{revision1}$B_T$} of the magnetic field in this setup measured with a Hall sensor at different positions along the beam axis. {\color{revision1}The steering magnet is set to a nominal field of 30~mT. An offset of 0.1~mT is added to all measurements} to make negative values visible on the logarithmic scale. The figure compares measurements at room temperature and with liquid nitrogen cooling. This shows that the 1~m long, 2-layer HTS tube shields most of the dipole field when it is in its superconducting state. We attribute the field distortions inside the shield to mechanical imperfections (which allow field to leak through the shield), as well as artifacts caused by background fields trapped inside the superconductor during cool-down. In addition, the ferromagnetic substrate of the superconductor causes field distortions both at room temperature and at cryogenic temperatures.
Figure \ref{fig:sbu_bvz} shows the same type of measurement for the 4.5~inch, 4-layer HTS and 4.5~inch, 45-layer HTS shields placed inside the Helmholtz coils setup. The distortions inside these shields are smaller. However, a significant fraction of the field leaks into the shields because the Helmholtz coils extend beyond the ends of these shields. This effect is stronger for the 4-layer HTS shield because its inner diameter is larger than the inner diameter of the 45-layer shield.
Figure \ref{fig:B_v_time} shows the field measured with a Hall sensor in the center of the 2-layer, 1~m long superconductor shield in the Van de Graaff setup as a function of time for nominal steering dipole fields of $B_{a}$~=~11~mT and $B_{a}$~=~45~mT. At the lower dipole setting, the field inside this shield is stable, while at the higher dipole setting it increases approximately logarithmically with time. As mentioned in Sec.~\ref{sec:scshielding}, such a time dependence above a certain threshold field is an expected behavior for HTS shields.
Figure \ref{fig:bnl_bvi} shows the magnetic field shielding performance of the 1~m long, 2-layer HTS shield in the Van de Graaff setup. The top panel presents the deflection of a 8.14~MeV $^{7}_{3}\mbox{Li}^{3+}$ beam as a function of the steering dipole field $B_{a}$, both at room temperature (superconductor does not shield field) and at liquid nitrogen temperature (superconductor shields field). The superconducting shield reduces the beam deflection by about 94\% (see \cite{kyle_conference} for a comparison of our $^{7}_{3}\mbox{Li}^{3+}$ and $^{16}_{8}\mbox{O}^{3+}$ results). The bottom panel shows the field measured with a Hall sensor in the center of the shield as a function of steering dipole fields. The open markers indicate field measurements that show the previously mentioned increase over time, i.e. measurements for which the mean reading {\color{revision1}in the last quarter is more than two standard deviations higher than the mean} reading in the first quarter of a ten minute measurement.
% The non-linear behavior of the room temperature measurements at high deflections is caused by the beam spot hitting the edge of the fluorescent screen.
Figure \ref{fig:sc_shield_mri} shows the shielding performance of our 45-layer HTS shield prototype in the MRI magnet in linear scale (top) and logarithmic scale (bottom). The open markers indicate field measurements showing an increase over time. The prototype shields more than 99\% of an external field up to 0.45~T, and 95\% up to 0.50~T external field.

% FERROMAGNET
\subsection{Ferromagnetic shells for magnetic field cloaking}
\label{sec:fmtuning}

\begin{figure}
	\centering
    	\includegraphics[width = 0.75\textwidth]{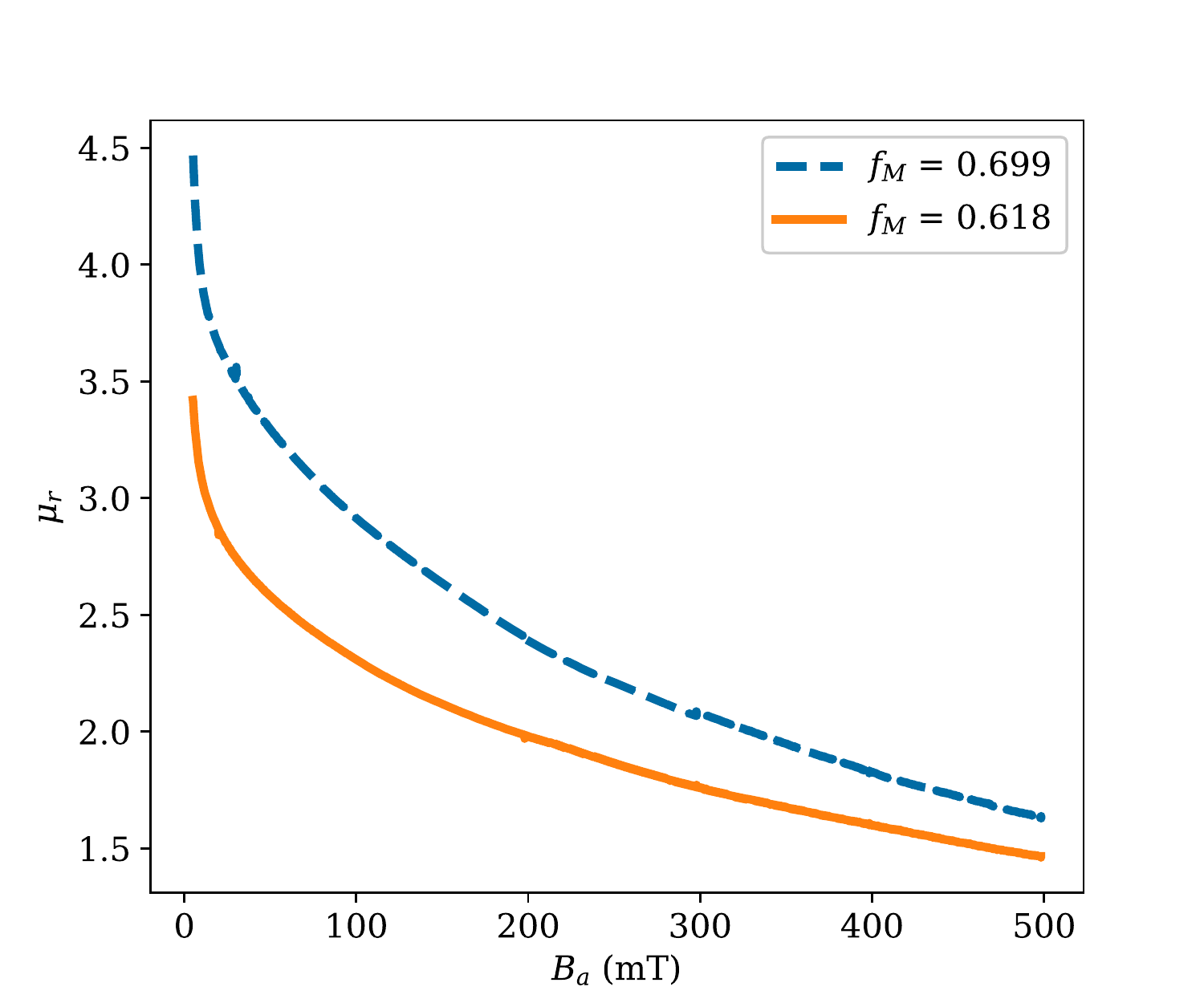}
    	\caption{Relative magnetic permeability $\mu_r$ of 4.5~inch {\color{revision1}long} epoxy / steel powder cylinders (with different fractional masses $f_M$ of steel powder in the mixture) as a function of the applied field $B_{a}$ measured in the MRI magnet.}
    	\label{fig:fm_mu_b}
\end{figure}

Figure \ref{fig:fm_mu_b} shows the magnetic permeability of two 4.5~inch long ferromagnetic epoxy / steel powder cylinders with different fractional masses $f_M$ of steel powder in the mixture at room temperature in the field of the MRI magnet as a function of the applied field $B_a$. The fractional mass $f_M$ is the mass of the steel powder used in the mixture divided by the mass of steel power plus epoxy. We determine the magnetic permeability from 
\begin{eqnarray}
    B_{T}(r < R_1) = \frac{4 \mu_r R_2^2}{(\mu_r +1)^2  R_2^2 - (\mu_r-1)^2 R_1^2} B_{a},   
    \label{eqn:findmu}
\end{eqnarray}
where {\color{revision1}$B_{T}$} and $B_{a}$ are the magnetic field measured in the center of the cylinder and the nominal MRI field, and $R_1$ and $R_2$ are the inner and outer radius of the ferromagnetic cylinder \cite{zangwill2012}. The magnetic permeability increases with higher $f_M$ and decreases with increasing applied field without saturating completely within the tested field range.

%\begin{figure}
%	\centering
%    	\includegraphics[width = 0.75\textwidth]{figures/ferromagnet/permeability_vs_fm_sbu}
%    	\caption{Magnetic permeability $\mu_r$ of various ferromagnetic epoxy / stainless steel powder cylinders at liquid nitrogen temperature and an external magnetic field of 40~mT as a function of fractional mass $f_M$ of these cylinders.}
%    	\label{fig:fm_mu_fmass}
%\end{figure}
%
%The magnetic permeability of our ferromagnetic cylinders is smaller at liquid nitrogen temperature than at room temperature, which we attribute to an increase in the anisotropy constant~\cite{oxley2009}. Figure \ref{fig:fm_mu_fmass} summarizes the relative magnetic permeability measured at liquid nitrogen temperature with the Helmholtz coil setup at an external field of 40~mT for various fractions $f_{M}$ of stainless steel powder in the epoxy mixture. This plot allows to select $f_M$ to fabricate a ferromagnetic cylinder of desired $\mu_r$.

\begin{figure}
	\centering
    	\includegraphics[width = 0.75\textwidth]{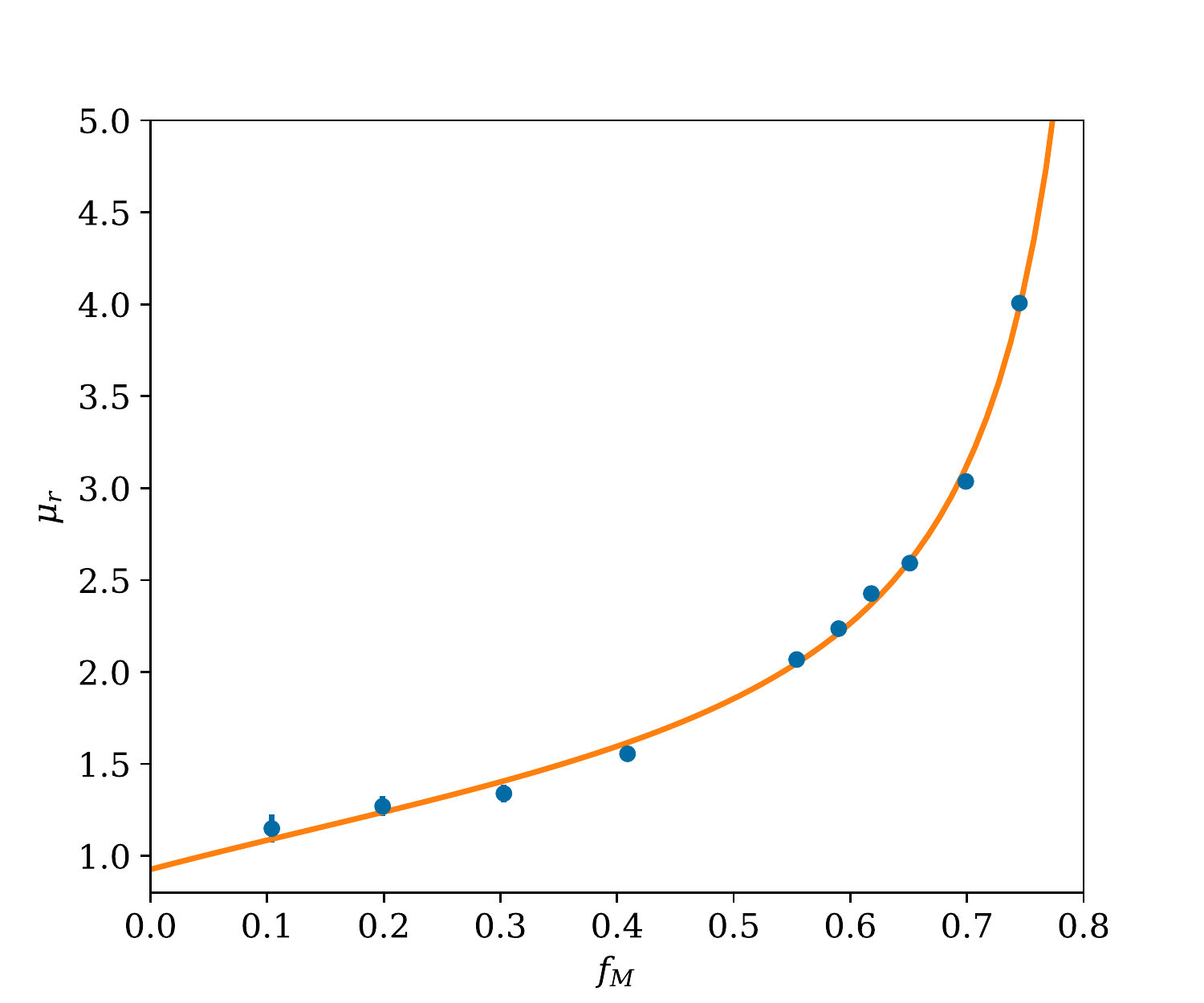}
    	\caption{Magnetic permeability $\mu_r$ of various ferromagnetic epoxy / stainless steel powder cylinders at liquid nitrogen temperature and an external magnetic field of 40~mT as a function of fractional mass $f_M$ of these cylinders. The line indicates a fit of Eqn.~\ref{eqn:calcmu} to the measurements with $\chi^2$/DOF~=~2.4 for the fit parameters listed in Table~\ref{tab:fit_mur_fm}.}
    	\label{fig:fm_mu_fmass}
\end{figure}

Figure \ref{fig:fm_mu_fmass} summarizes the relative magnetic permeability measured at liquid nitrogen temperature with the Helmholtz coil setup at an applied field of 40~mT for 4.5~inch long ferromagnetic cylinders of various $f_{M}$. We empirically find that the function
\begin{eqnarray}
    \mu_r (f_M) \; = \; \frac{p0}{\tan(p1 \, \cdot \, f_M \, + \, p2)} \, + \, p3   
    \label{eqn:calcmu}
\end{eqnarray}
describes the relation between $\mu_r$ and $f_M$ in these cylinders with the parameters listed in table \ref{tab:fit_mur_fm} and a $\chi^2/\mbox{DOF}$ of 2.4. This plot and function allow to select $f_M$ to fabricate a ferromagnetic cylinder of desired $\mu_r$. The magnetic permeability at liquid nitrogen temperature is smaller than at room temperature, which we attribute to an increase in the anisotropy constant~\cite{oxley2009}. This needs to be accounted for when selecting $f_{M}$.

\begin{table}[h!]
\centering
\begin{tabular}{|c c c c|} 
 \hline
 p0 & p1 & p2 & p3 \\ [0.5ex] 
 0.70 $\pm$ 0.05 & -2.2 $\pm$ 0.4 & 1.9 $\pm$ 0.4 & 1.1 $\pm$ 0.2 \\ [1ex] 
 %0.70(5) & -2.2(4) & 1.9(4) & 1.1(2) \\ [1ex] 
 \hline
\end{tabular}
\caption{Fit parameters for Eqn.~\ref{eqn:calcmu} and the line shown in Fig.~\ref{fig:fm_mu_fmass}.}
\label{tab:fit_mur_fm}
\end{table}

%CLOAKING
\subsection{Magnetic field cloaking}

\begin{figure}
	\centering
    	\includegraphics[width = 0.75\textwidth]{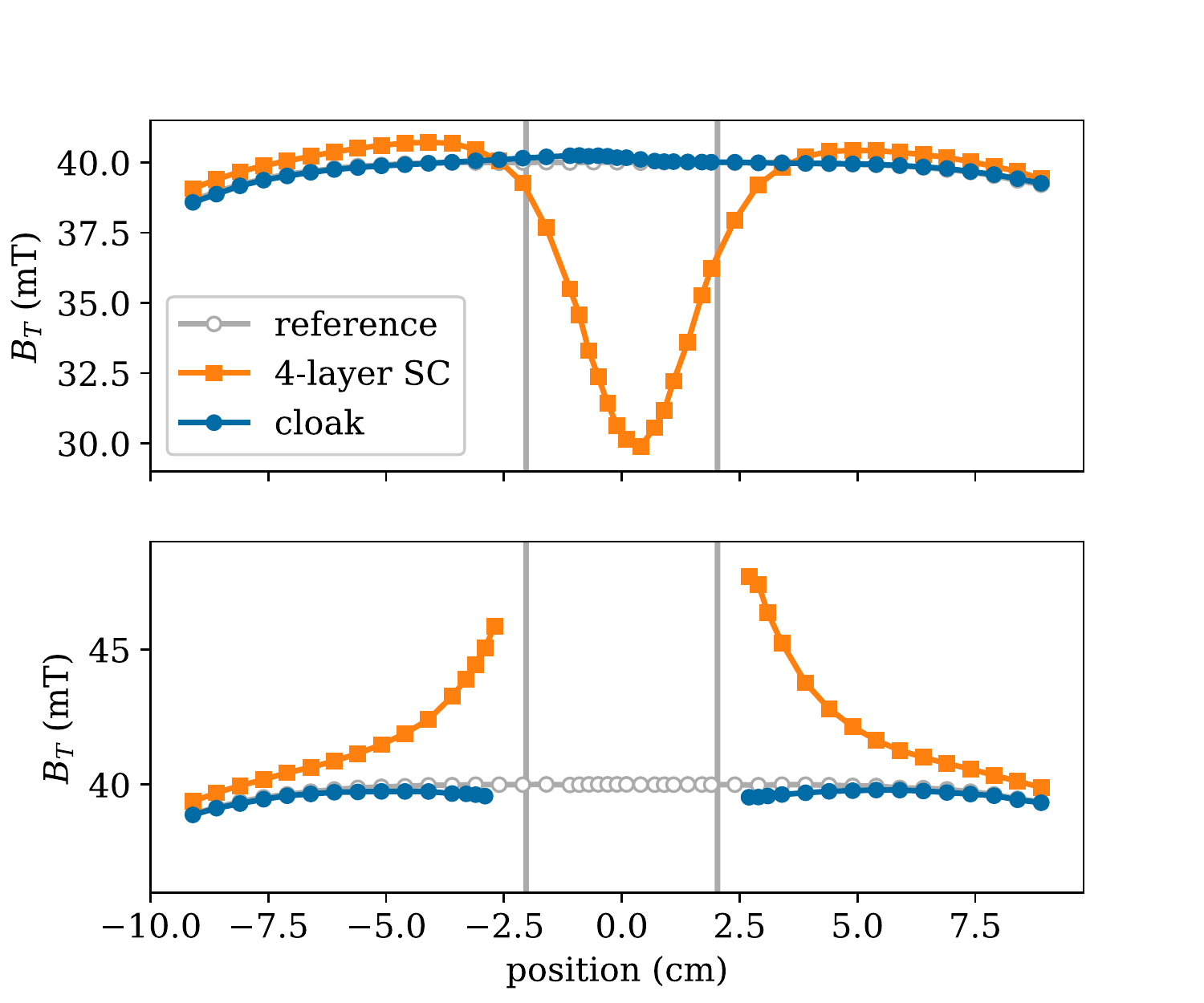}
    	\includegraphics[width = 0.115\textwidth]{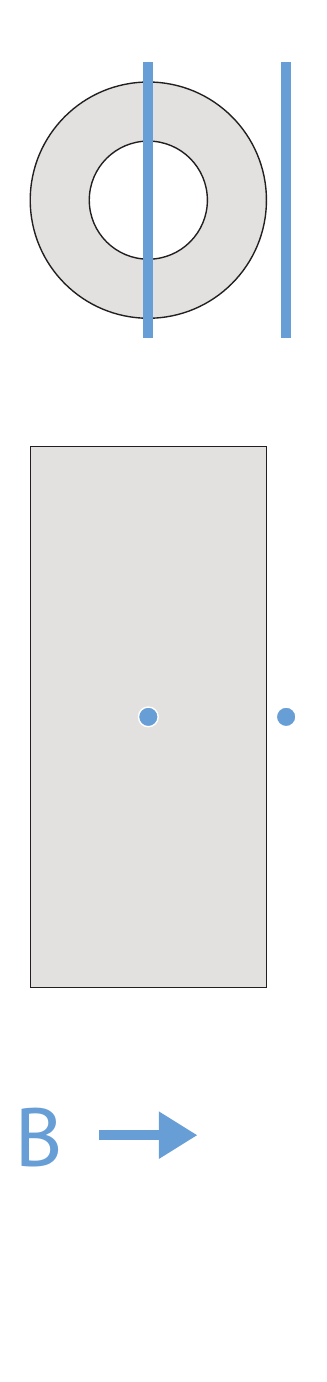}
    	\caption{Magnetic flux {\color{revision1}$B_{T}$} in the direction of the Helmholtz coils axis at positions along measurement lines at 1~cm distance from the cloak (4.5~inch long, 4 layer HTS and ferromagnet with {\color{revision1}$\mu_r$~=~2.43(4)}, top panel) and across its center (bottom panel). The plots also show the field for the coils with only the superconducting cylinder (4-layer SC) and the coils itself (reference) measured at the same positions. Vertical lines indicate the cloak dimensions.}
    	\label{fig:cloak_bvx_1}
\end{figure}

\begin{figure}
	\centering
    	\includegraphics[width = 0.75\textwidth]{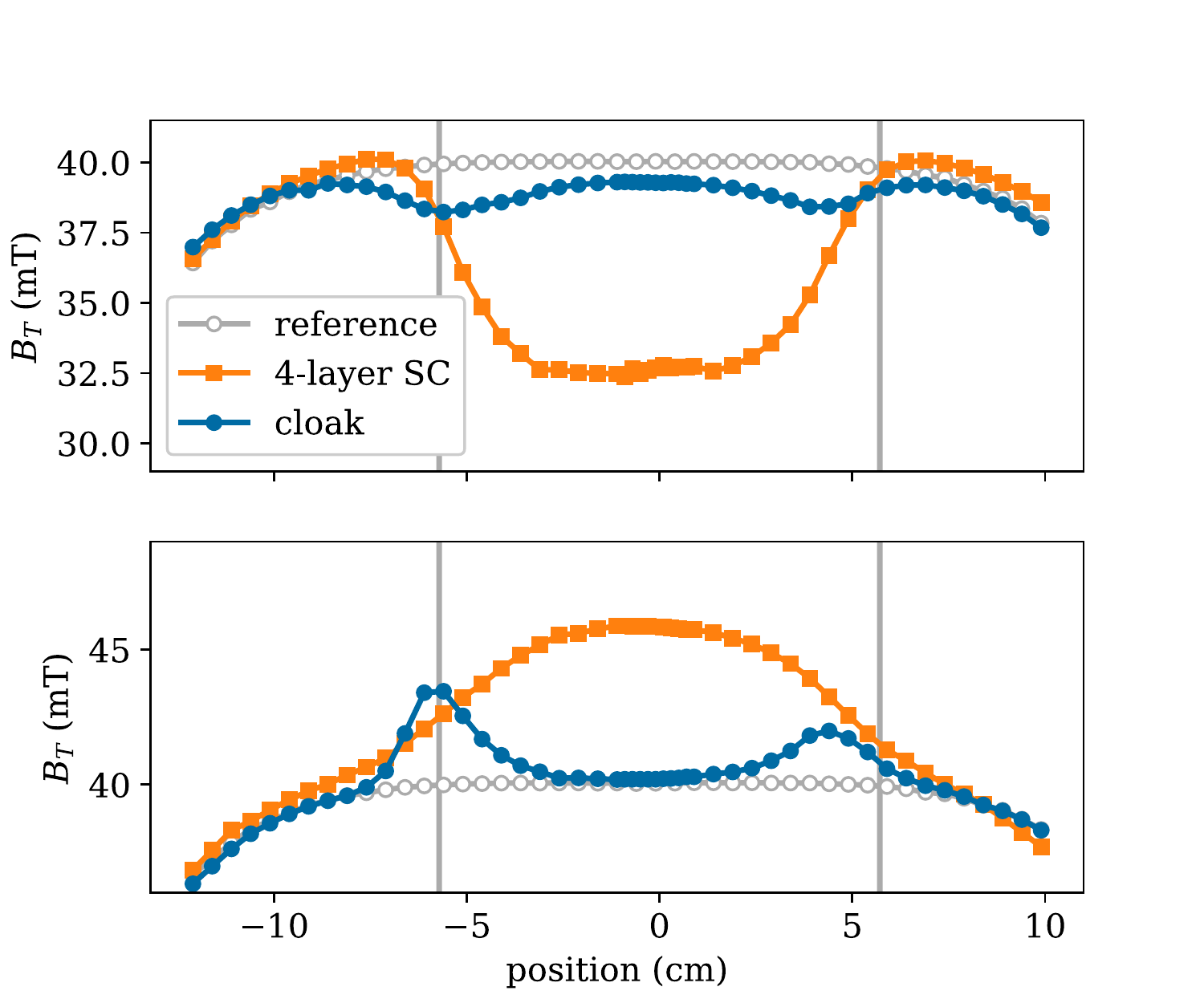}
    	\includegraphics[width = 0.115\textwidth]{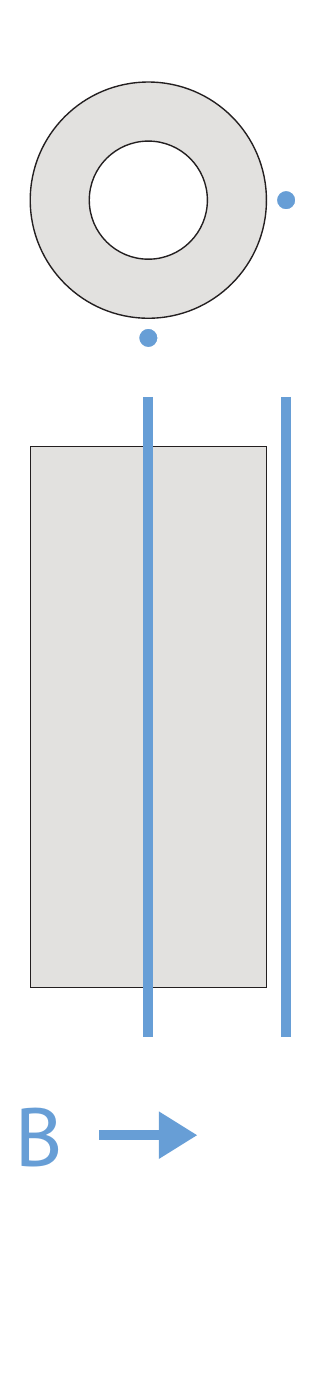}
    	\caption{Magnetic flux {\color{revision1}$B_{T}$} in the direction of the Helmholtz coils axis at positions along measurement lines at 1~cm distance in front of (top figure) and next to (bottom figure) the cloak (4.5~inch long, 4 layer HTS and ferromagnet with {\color{revision1}$\mu_r$~=~2.43(4)}). The plots also show the field for the coils with only the superconducting cylinder (4-layer SC) and the coils itself (reference) measured at the same positions. Vertical lines indicate the cloak dimensions.}
    	\label{fig:cloak_bvz_1}
\end{figure}

Based on the diameters of our 4.5~inch long cloak prototypes, Eqn.~\ref{eqn:permeability} predicts perfect cloaking with a ferromagnetic shell {\color{revision1} that has a relative magnetic permeability of 2.56(2)}. Figures \ref{fig:cloak_bvx_1} and \ref{fig:cloak_bvz_1} show maps of the Helmholtz coils magnetic field at a nominal applied field of 40~mT for the coils itself, the coils with a 4.5~inch long, 4-layer HTS shield, and the coils with a full cloak (also using the 4.5~inch long, 4-layer HTS shield).  {\color{revision1}The ferromagnetic shell used for this cloak has a $\mu_r$ of 2.43(4) (which is 5\% below the theoretically ideal value of 2.56(2))} at liquid nitrogen temperature and at 40~mT applied field. These figures demonstrate that the cloak significantly reduces the field disturbances that the superconducting cylinder alone would cause. Because the field of the Helmholtz coils extend beyond the ends of the cloak, the figures  show fringe effects near the ends of the cloak. 

%The higher the field, the more field leaks through. Ref.~\cite{Denis2007} discusses an exponential relationship of the leaked field at the extremities of the tube. Beyond the fringe area, the field profile is relatively flat.

\begin{figure}
	\centering
    	\includegraphics[width = 0.75\textwidth]{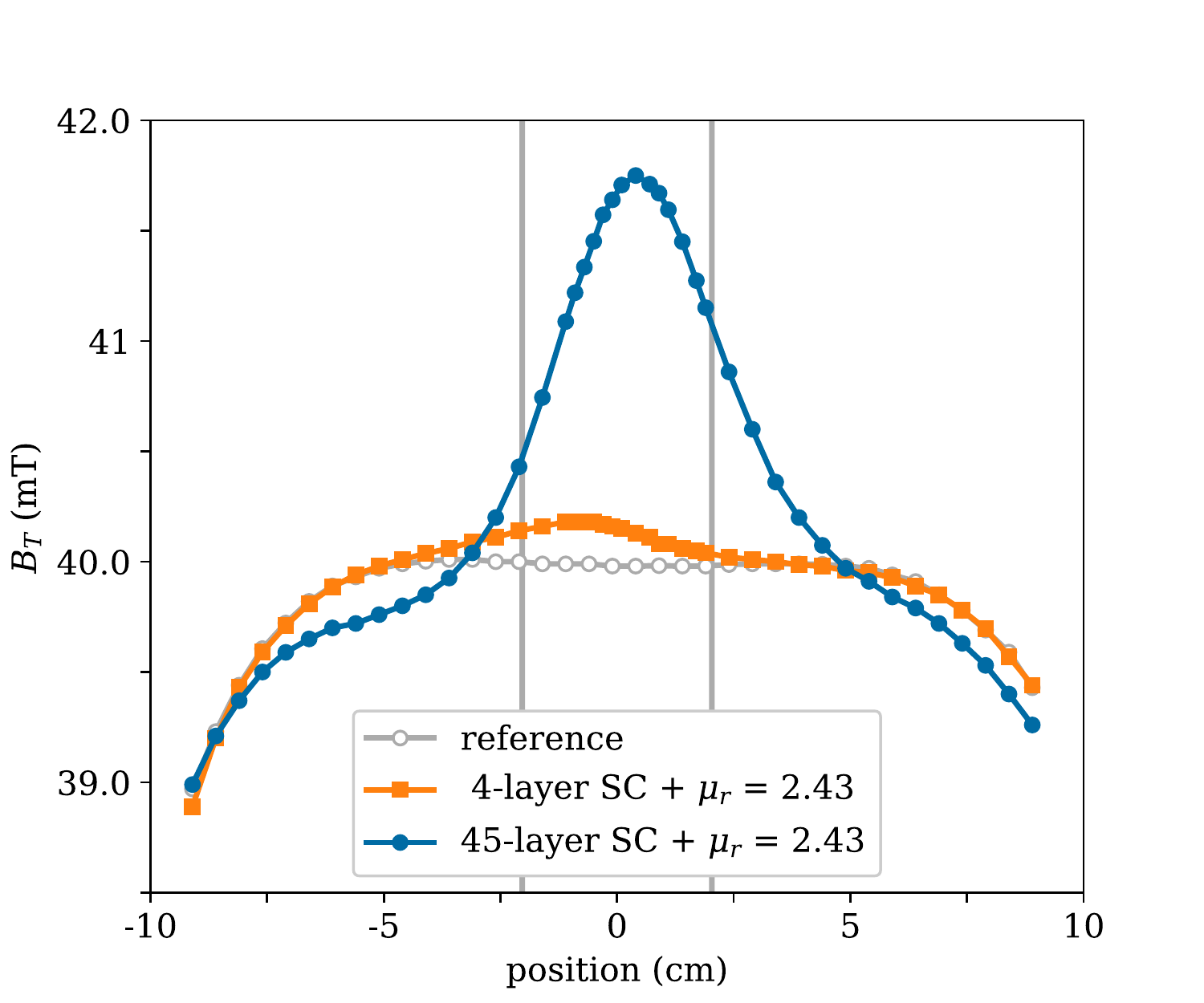}
    	\includegraphics[width = 0.115\textwidth]{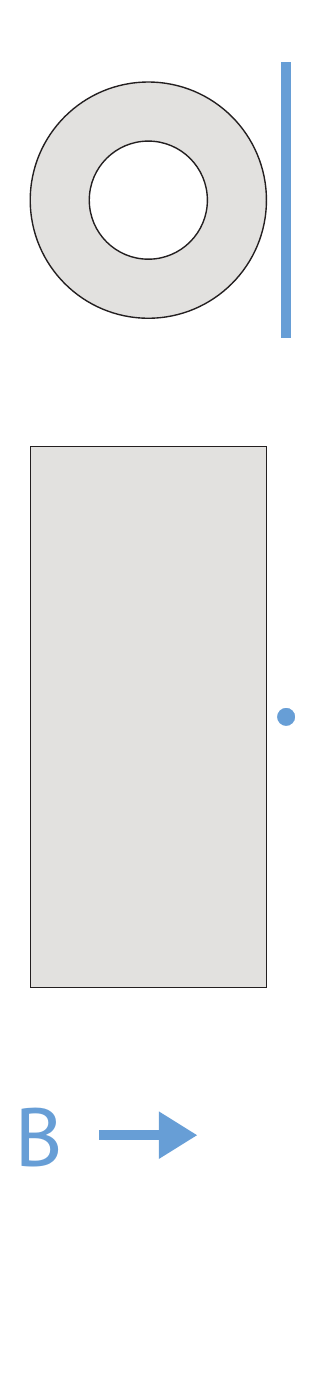}
    	\caption{Magnetic flux {\color{revision1}$B_{T}$} in the direction of the Helmholtz coils axis at positions along a measurement line (the same positions as the top of Fig. \ref{fig:cloak_bvx_1}) at 1~cm distance from the cloak with the 4.5~inch long, four-layer and 4.5~inch long, 45-layer superconductor combined with the same ferromagnet shell ({\color{revision1}$\mu_r$~=~2.43(4)}). Vertical lines indicate the cloak dimensions.}
    	\label{fig:cloak_diff_sc}
\end{figure}

\begin{figure}
	\centering
    	\includegraphics[width = 0.75\textwidth]{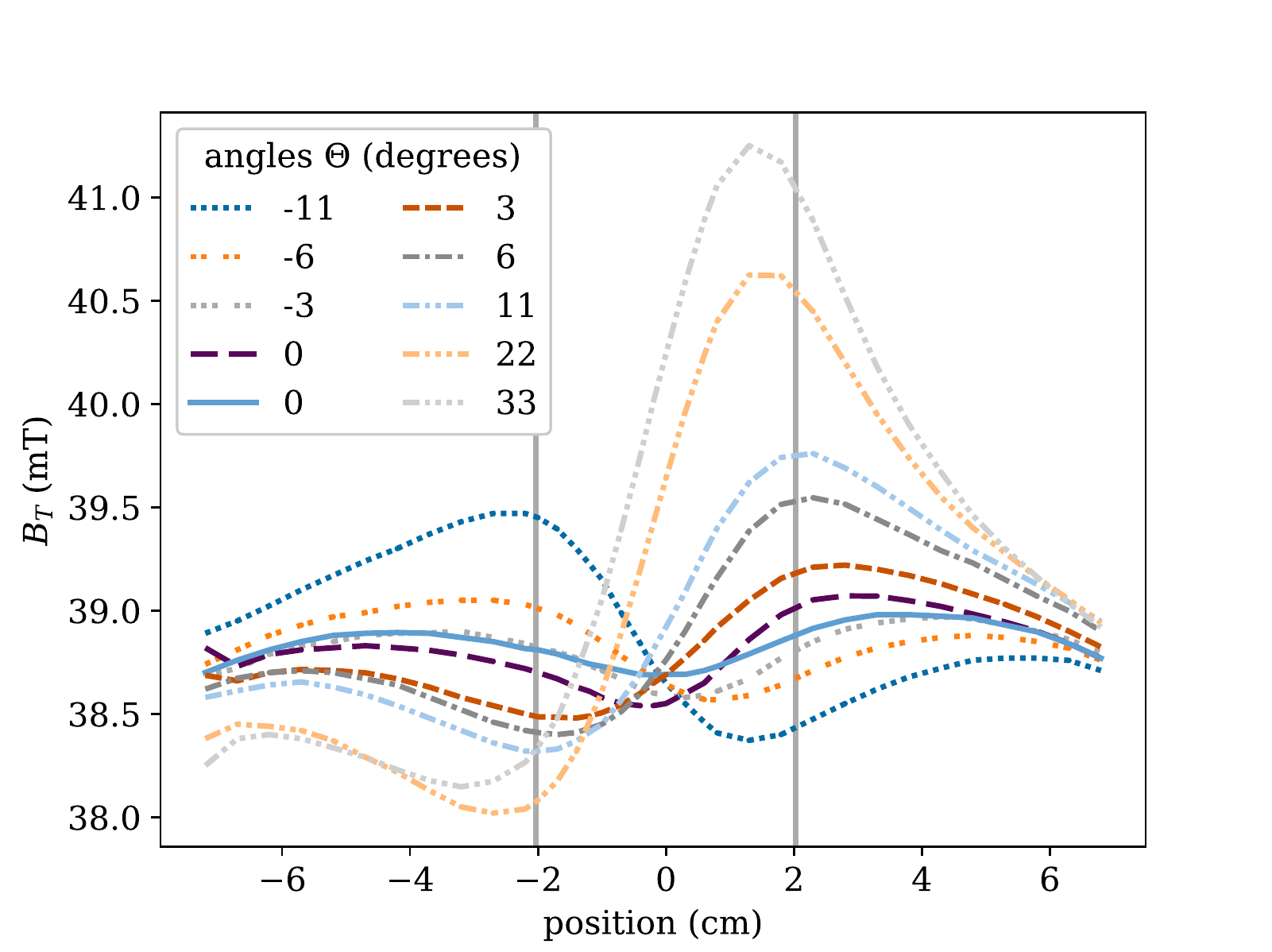}
    	\includegraphics[width = 0.24\textwidth]{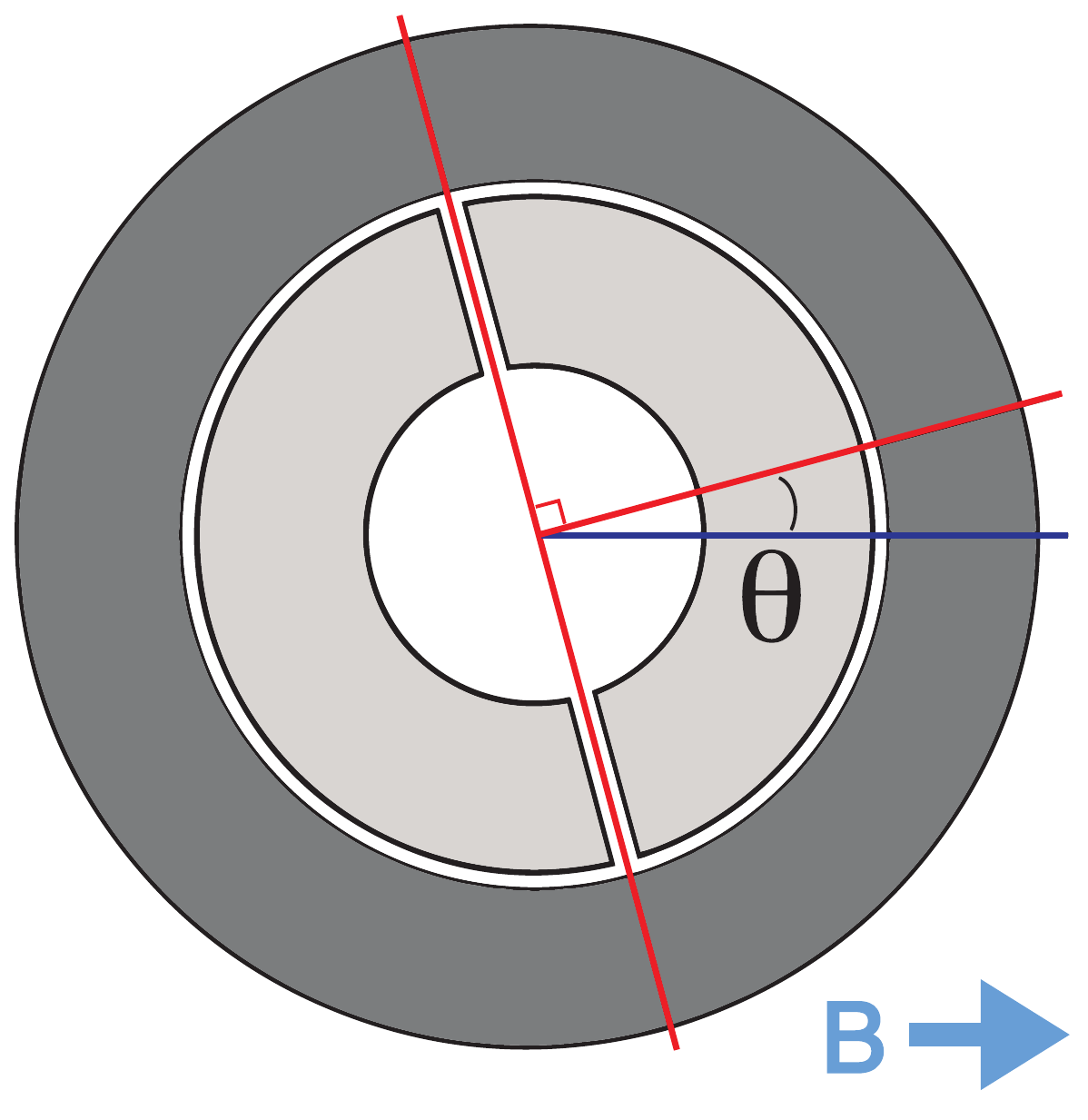}
    	\caption{Magnetic flux {\color{revision1}$B_{T}$} in the direction of the Helmholtz coils axis at positions along a measurement line (the same positions as the top of Fig. \ref{fig:cloak_bvx_1}) at 1~cm distance from the cloak (4.5~inch long, 45 layer HTS and ferromagnet with {\color{revision1}$\mu_r$~=~2.43(4)}) for different angles between the gap separating the two halves of the HTS shield and the magnetic field. Vertical lines indicate the cloak dimensions.}
    	\label{fig:cloak_angle}
\end{figure}

Shielding and cloaking higher fields requires switching to the 4.5~inch long, 45-layer HTS cylinder. Due to imperfections in the fabrication, this cylinder is not perfectly cylindrical and leaves a gap of about 0.1~inch when inserted into the ferromagnetic shell. Using this superconductor with the ferromagnet with {\color{revision1}$\mu_r$ = 2.43(4)} yields increased field disturbances around the cloak (see Fig. \ref{fig:cloak_diff_sc}). We find that we can compensate for some of these effects by selecting a lower permeability ferromagnet, or using this ferromagnet at higher fields, which effectively reduces its $\mu_r$. In addition, the 45-layer HTS shield has a gap where the two half-shells connect. Figure \ref{fig:cloak_angle} shows the field distortions near the cloak caused by different alignments of the gap with respect to the applied field. %The latter demands careful alignment of the superconductor with respect to the applied field.

\begin{figure}
    	\includegraphics[width = 0.75\textwidth]{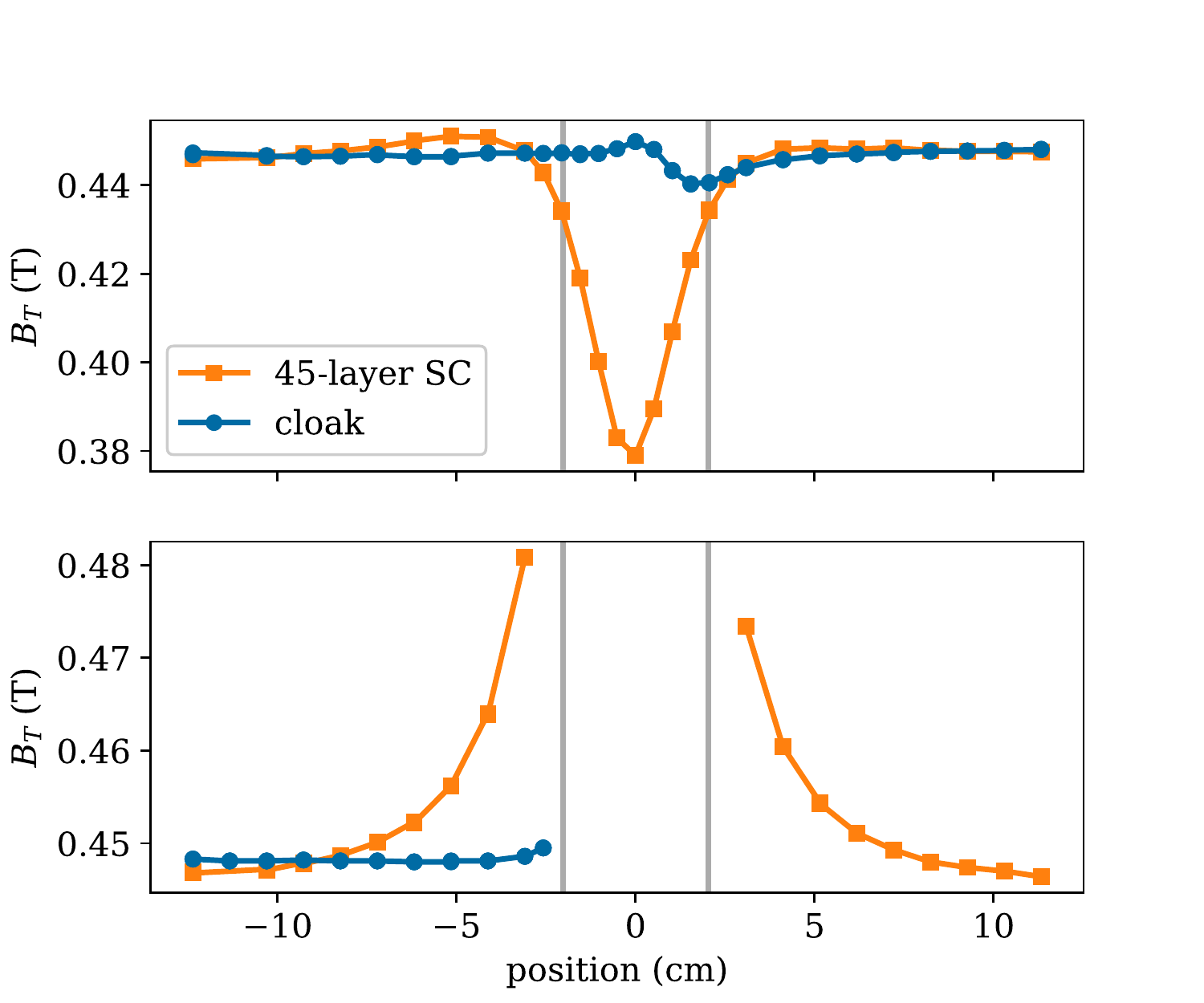}
	\includegraphics[width = 0.115\textwidth]{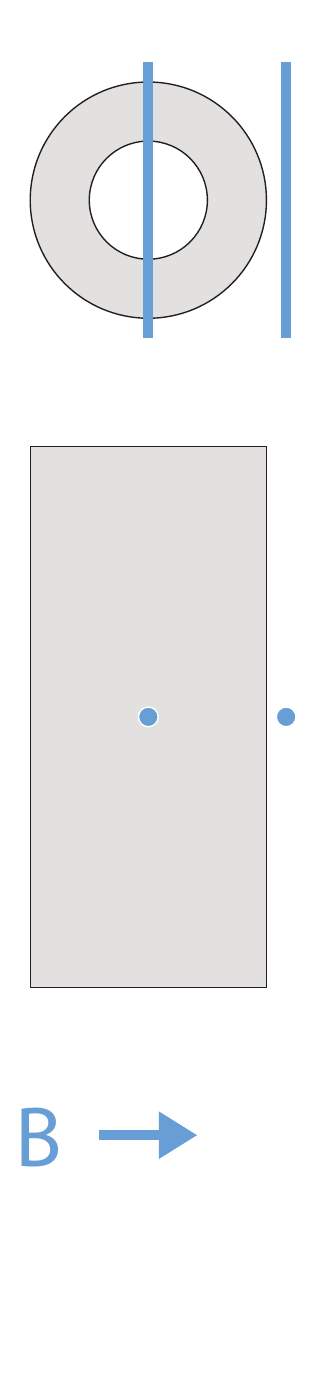}
    	\caption{Magnetic flux in the direction of the MRI axis at positions along measurement lines at 1~cm distance from the cloak (top) and across its center (bottom). The plots also show the field if only the superconducting cylinder is in the magnet measured at the same positions. Vertical lines indicate the cloak dimensions.}
    	\label{fig:cloak_mri_1d_front}
\end{figure}

\begin{figure}
	\includegraphics[width = 0.75\textwidth]{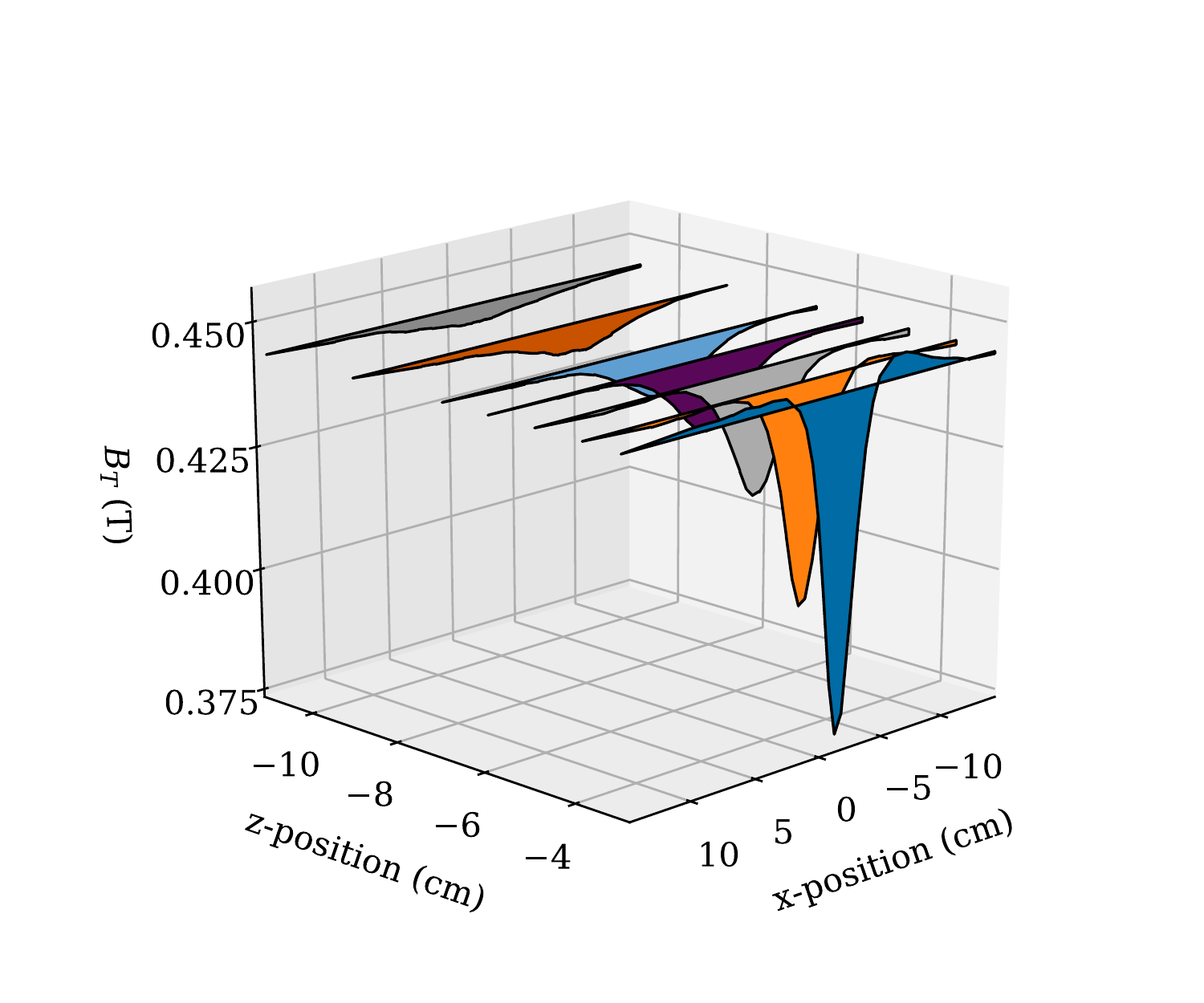}
	\includegraphics[width = 0.24\textwidth]{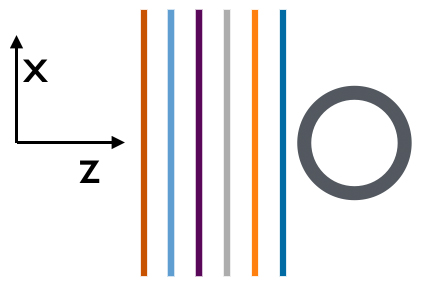}
    	\includegraphics[width = 0.75\textwidth]{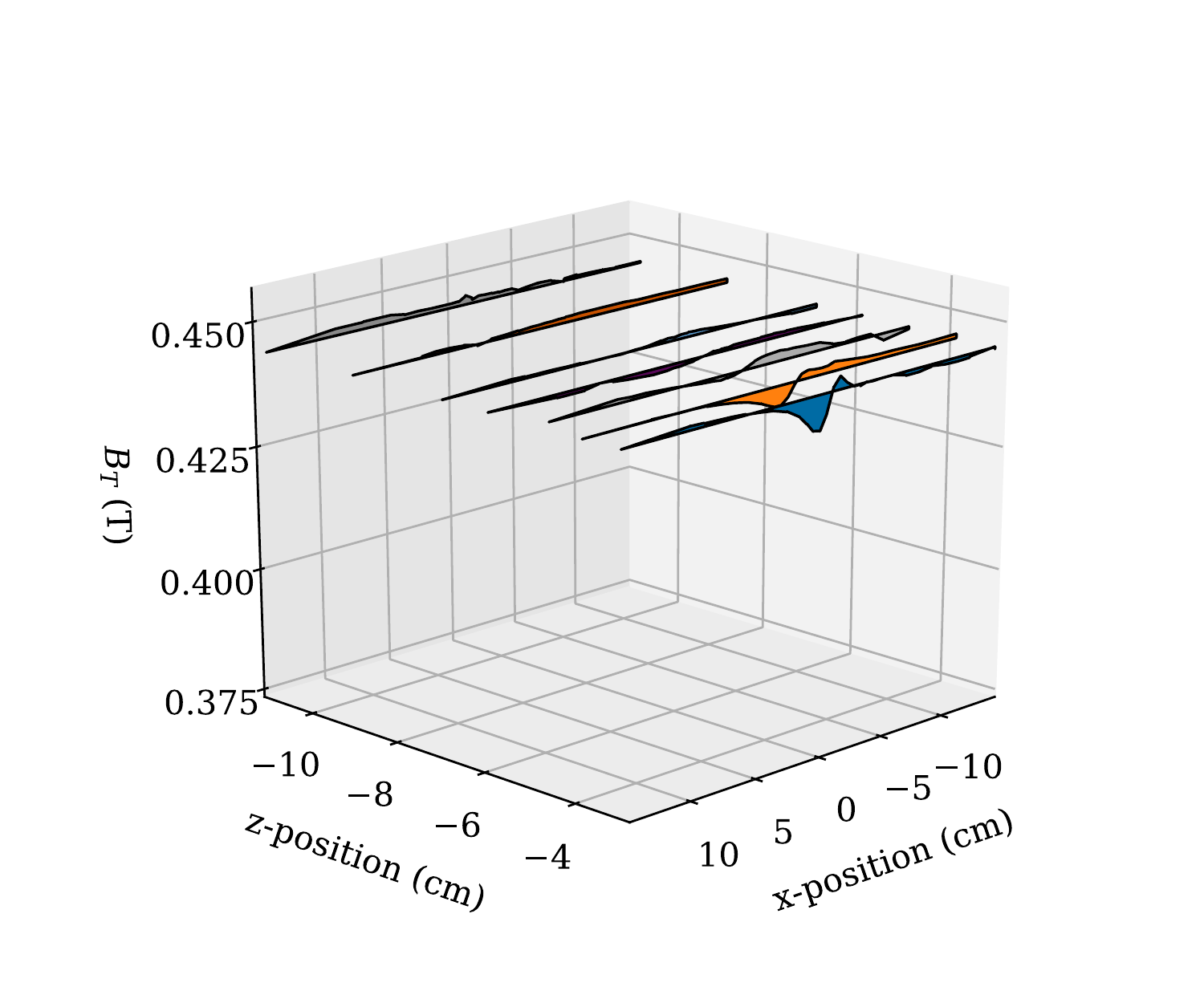}
	\includegraphics[width = 0.24\textwidth]{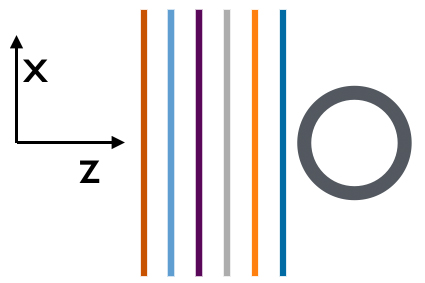}
    	\caption{Magnetic flux {\color{revision1}$B_{T}$} in the direction of the MRI axis at positions along measurement lines at various distances from the 4.5~inch long, 45-layer HTS shield (top panel) and the cloak, i.e. the same superconductor shield surrounded by a ferromagnetic shell with {\color{revision1}$\mu_r$~=~2.43(4)} (bottom panel). The nominal field of the MRI magnet is 0.45~T.}
    	\label{fig:cloak_mri_3d}
\end{figure}

To demonstrate magnetic field cloaking in the MRI magnet, we map the magnetic flux {\color{revision1}$B_{T}$} in the direction of the MRI axis near the 4.5~inch long, 45-layer HTS shield and a full cloak (same HTS shield and a ferromagnetic shell with {\color{revision1}$\mu_r$~=~2.43(4)}). Figure \ref{fig:cloak_mri_1d_front} shows that at 0.45~T nominal MRI field, the HTS shield causes a field distortion with an amplitude of 16\% of the field at 1~cm distance, while the distortion near the cloak is only 2\% of the applied field. The shape of this distortion hints at a misalignment of the superconductor with respect to the MRI field (see Fig.~\ref{fig:cloak_angle}). At 0.5~T nominal MRI field, the distortion amplitude is 13\% of the field near the the HTS shield alone and 2\% near the cloak. Figure \ref{fig:cloak_mri_3d} illustrates the field distortions caused by the HTS shield at up to 10~cm from the superconductor and that the cloak significantly mitigates these distortions.

%______________________________________________________________________________
\section{Conclusion}
\label{sec:conclusion}
Our tests demonstrate that a magnetic field cloak is a viable option to shield charged particle beams from transverse magnetic fields of up to at least 0.5~T. This allows for new designs of collider experiments and enables, for example, the use of dipole magnets in the forward regions of an EIC detector to improve the momentum measurement of charged final state particles at angles close to the beam line. The number of HTS layers used affects the maximum shielded field. At the same time, cooling the cloak with liquid helium instead of liquid nitrogen would significantly increase the maximum shielded field for a given number of layers, or would reduce the number of layers needed to shield a given field. The cloaking performance can be maximized by carefully tuning the permeability of the ferromagnet and precisely fabricating the HTS cylinder. Moreover, ensuring proper alignment of the HTS shield w.r.t the magnetic field or using a gap-free HTS cylinder (e.g. by evaporating the superconductor directly onto a core) would help to minimize field distortions around the cloak. Fringe field effects at the ends of the cloak can be mitigated by extending the cloak beyond the field. The design parameters, fabrication procedures, and limitations of a magnetic field cloak established in this paper pave the way to realize such a device. This, in turn, opens up new possibilities to design future collider facilities and experiments.

%______________________________________________________________________________
\section{Acknowledgements}

This work was supported by the DOE Office of Nuclear Physics' Electron-Ion Collider Detector R\&D initiative administered by Brookhaven National Laboratory (Contract No. DE-AC02-98CH10886). Special thanks go to the program's advisory committee for their helpful guidance and facilitation. We also thank B. Parker, R. Gupta, V. Ptitsyn, Y. Goto, K. Boyle, I. Nakagawa, and J. Seele for valuable discussions, as well as American Superconductors for making their 46~mm wide superconductor wire insert available to us. In addition, we gratefully acknowledge the support of the staff of the Tandem Van de Graaff Facility at Brookhaven National Laboratory, the 4 Tesla Magnet Facility at Argonne National Laboratory, and the Stony Brook Physics Department machine shop. We thank S. Karthas, R. Lefferts, A. Lipski, and I. Yoon for all their valued help. A significant portion of this research project was carried out by undergraduate students, and we thank
D. Aviles,
G. Bello Portmann,
D. Bhatti,
I. Bromberg,
R. Bruce,
J. Chang,
B. Chase,
E. Jiang,
P. Karpov,
Y. Ko,
Y. Kulinich,
R. Losacco,
E. Michael,
J. Nam,
H. Powers,
V. Shetna,
S. Thompson,
H. Van Nieuwenhuizen, and
N. Ward for their contributions. Finally, we thank A. Chhugani, and the Simons Summer Research Program for enabling her to participate in this project as a high school student.

\clearpage
\section*{References}
\bibliography{magcloak_references}

\clearpage
%\appendix
%\section{Other images not included in current draft}
%\newpage
%
%\begin{figure}
%	\centering
%    	\includegraphics[height = 0.3\textheight]{figures_v0/SCCriticalField}
%    	\caption{Characterize performance of HTS tubes by measuring leakage field B$_i$ as a function of the applied field B$_i$}
%    	\label{fig:sc_experiment}
%\end{figure}
%
%\begin{figure}
%  	\centering
% 	\includegraphics[height = 0.4\textheight]{figures_v0/photo_die_mandrel}
%	\caption{Photo of die and mandrel setup for laminating superconductor layers.}
%	\label{fig:die_mandrel}
%\end{figure}
%
%\begin{figure}
%    	\centering
%    	\includegraphics[width=0.75\textwidth]{figures_v0/photo_bnl_setup}
%    	\caption{Photo of beam test prototype.}
%    	\label{fig:vdgshield}
%\end{figure}
%
%\begin{figure}
%    	\centering
%    	\input{figures_v0/extrapolating_performance4.pgf}
%    	\caption{B$_i$ vs B$_o$ measurements for 
%        superconducting cylinders made from 1-4 layers of the
%        46~mm wide superconductor tape.}
%    	\label{fig:sc_shield_hh}
%\end{figure}
%
%
%
%\begin{figure}
% 	\centering
%    	\includegraphics[height = 0.4\textheight]{figures_v0/sc_shielding_hh_vs_dipole}
%    	\caption{Plot: Bi vs B0 same prototype in Helmholtz and Dipole.}
%    	\label{fig:sc_shield_hh_vs_dipole}
%\end{figure}

\end{document}